\documentclass[default,fleqn,iicol]{sn-jnl}

\usepackage{amsmath}

\jyear{2022}%

\theoremstyle{thmstyleone}%
\theoremstyle{thmstyletwo}%

\theoremstyle{thmstylethree}%

\begin{document}

\title[Article Title]{Design study for a 500 MeV positron beam at the Mainz Microtron MAMI}


\author{\fnm{H.} \sur{Backe}}  
\author{\fnm{W.} \sur{Lauth}}   
\author{\fnm{P.} \sur{Drexler}}
\author{\fnm{P.} \sur{Heil}}
\author{\fnm{P.} \sur{Klag}}
\author{\fnm{B.} \sur{Ledroit}}
\author{\fnm{F.} \sur{Stieler}}


\affil{\orgdiv{Institut für Kernphysik}, \orgname{Johannes Gutenberg Universität},  \city{Mainz}, \postcode{55028},  \country{Germany}}



\abstract{A design study has been performed for a positron beam with an energy of 500 MeV to be realized at the applied physics area of the Mainz Microtron MAMI. Positrons will be created after pair conversion of bremsstrahlung, produced by the 855 MeV electron beam af MAMI in a tungsten converter target. From the two conceivable geometries (i) pair conversion in the bremsstrahlung converter target itself, and (ii) bremsstrahlung pair conversion in a separated lead foil, the former was considered in detail. Positrons will be energy selected within an outside open electron beam-line bending magnet, and bent back by an additional sector magnet. Magnetic focusing elements in between are designed to prepare in a well shielded positron target chamber about 6 m away from the target a beam with horizontal and vertical emittances of $\epsilon_v $ = 0.055 $\pi$ mm mrad (1 $\sigma$), and $\epsilon_h $ = 0.12 $\pi$ mm mrad (1 $\sigma$), respectively, for a 10 $\mu$m thick amorphous tungsten target and negligible momentum spread. At an accepted positron band width of 1 MeV, spots  are expected vertically with an angular spread of 0.064 mrad and a size of 5.0 mm (FWHM), and horizontally with an angular spread of 0.64 mrad and a size of 7.7 mm (FWHM). The positron yield amounts to 13.1 per second, 1 MeV positron energy band width, and 1 nA electron beam current.}




\maketitle

\section{Introduction}\label{sec1}

The channeling phenomenon of charged particles in single crystals has been applied in a number of areas. Well known is the particle steering in high energy physics, but considerable interest exists also in the construction of compact radiation sources in the MeV range and beyond, for an overview see e.g. Korol et al. \cite{Korol:14}. In these fields bent or periodically bent single crystals are required, the production of which is already an art in itself, and the understanding of the channeling process in such crystals is of utmost importance not only for electrons but in particular also for positrons. To study channeling in such crystals, high quality low emittance beams in the GeV range and below are required.

For electrons such a facility is the accelerator complex MAMI at the Institute for Nuclear Physics of the University of Mainz which supplies an electron beam with a maximum energy of 1.6~GeV and a beam current of up to 100 µA. Outstanding qualities of MAMI are the continuous beam with an excellent beam quality of some $\pi$ nm rad emittance, a very low energy spread of less than $10^{-4}$, as well as its extremely high reliability \cite{Jankowiak:06}. The electron beam is mainly used for nuclear physics experiments but the low divergence beam is well suited also for all kind of channeling experiments in thin bend crystals. Such experiments require a high quality beam with a divergence less than the Lindhard angle which is typically less than 0.3~mrad at 1~GeV. In the past years, a large number of investigations on channeling radiation \cite{Backe:08, Lauth:08, Backe:15, Backe:18B}, deflection of the electron beams in bent crystals \cite{Mazzolari:14, Bandiera:21} and the generation of radiation in periodically bent crystals \cite{Wistisen:14, Wistisen:17, Backe:13C} have been conducted at MAMI.

Extending such studies for positrons hampers on the fact that facilities for high quality positron beams are rare. There exists the DESY II test beam facility \cite{Diener:2019} with selectable momenta in the range from 1 to 6 ~GeV/c. Another is the DA$\Phi$NE beam test facility at INFN Frascati \cite{Mazzitelli:03} which delivers a pulsed positron beam in the energy range around 500 MeV. Outside Europe we mention the End Station A Test Beam at SLAC \cite{Woods:2005}.

The final aim of the current design study is to construct a 500 MeV positron beam line with an outstanding emittance in one dimension, the vertical coordinate, allowing for a beam divergence of less than 100 $\mu$rad. To achieve this goal, the 855 MAMI beam is focused on a self-converter tungsten foil in which bremsstrahlung photons and positrons via pair conversion are produced in one and the same target, utilizing this way the high beam quality of MAMI. With such a beam the transverse energy in a crystal potential pocket, with a typical depth of 23 eV, could be defined at channeling with an energy spread in the order of eV.

The paper is organized as follows. After this Introduction, in section \ref{Positron_creation} the positron production via bremsstrahlung and pair conversion is described. The key part of the paper is section \ref{ImagingBM1} in which the calculation of the transfer matrix for positrons after passage of an outside open beam line bending magnet is described. In order to verify the calculations in sections \ref{Positron_creation} and \ref{ImagingBM1}, test experiments were performed which are described in section \ref{experimentaltests}. In section \ref{raytracing} more or less standard ray trace calculations are described to transport the positron beam into the experimental area. The paper closes in sections \ref{discussion} and \ref{conclusions} with a discussion and conclusions, respectively.

\begin{figure}[]
\centering

   \includegraphics*[angle=0,scale=0.25,clip]{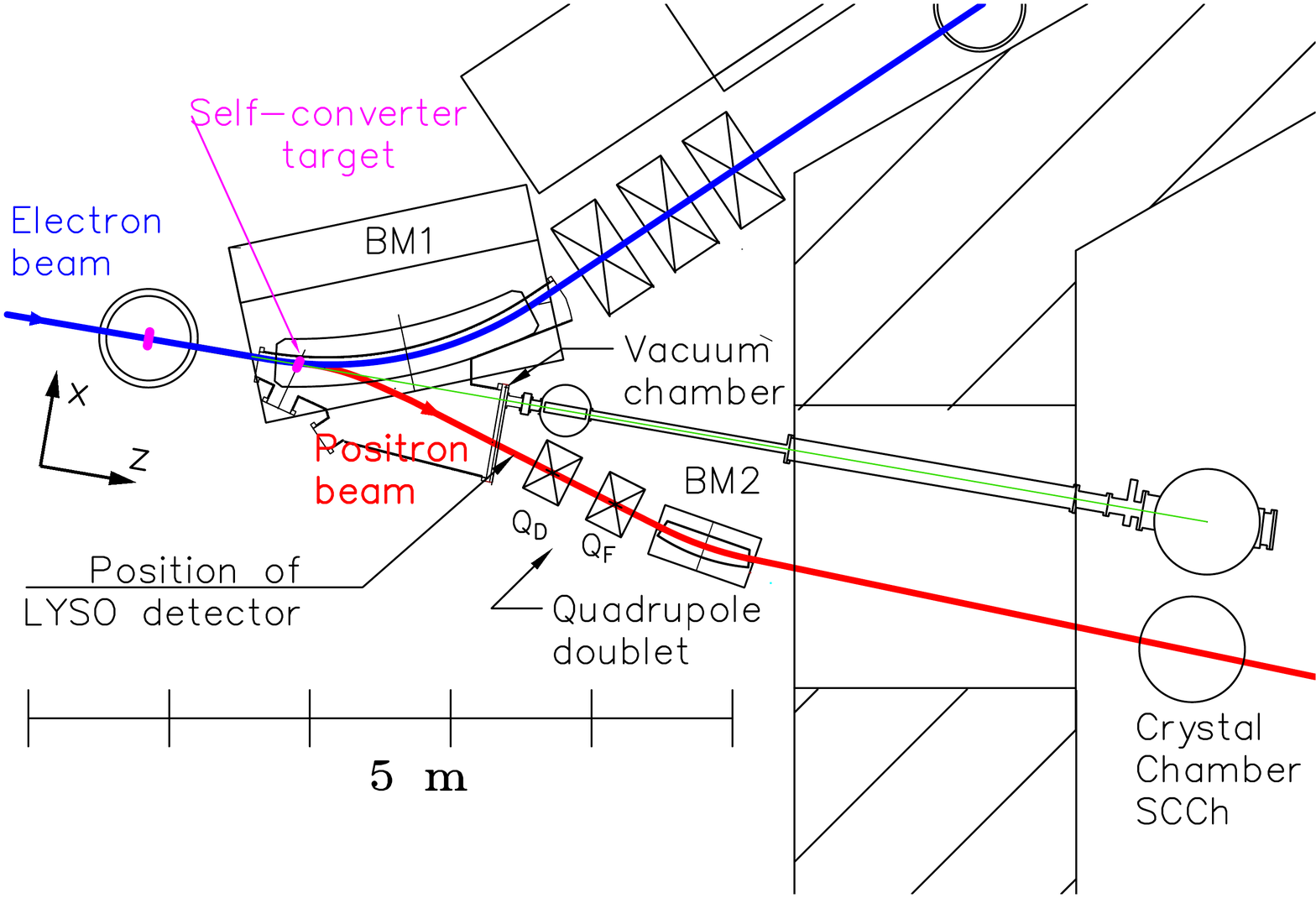}
\caption{
Sketch of the positron beam line in red at the X1 experimental area of MAMI. The displacement for the 500 MeV positron beam at the Single Crystal Chamber (SCCh) amounts to 1006 mm. Not shown are equipments behind the SCCh like a BM3 which deflects the positrons into a coincidence detector, and a photon detector for channeling or crystal undulator radiation in a distance of about 5-8 m from the SCCh.
}
\label{Exp-setup}  
\end{figure}

\section{Positron creation}
\label{Positron_creation}
\begin{figure}[tb]
\hspace{-1cm}
\includegraphics*[angle=0,scale=0.25,clip]{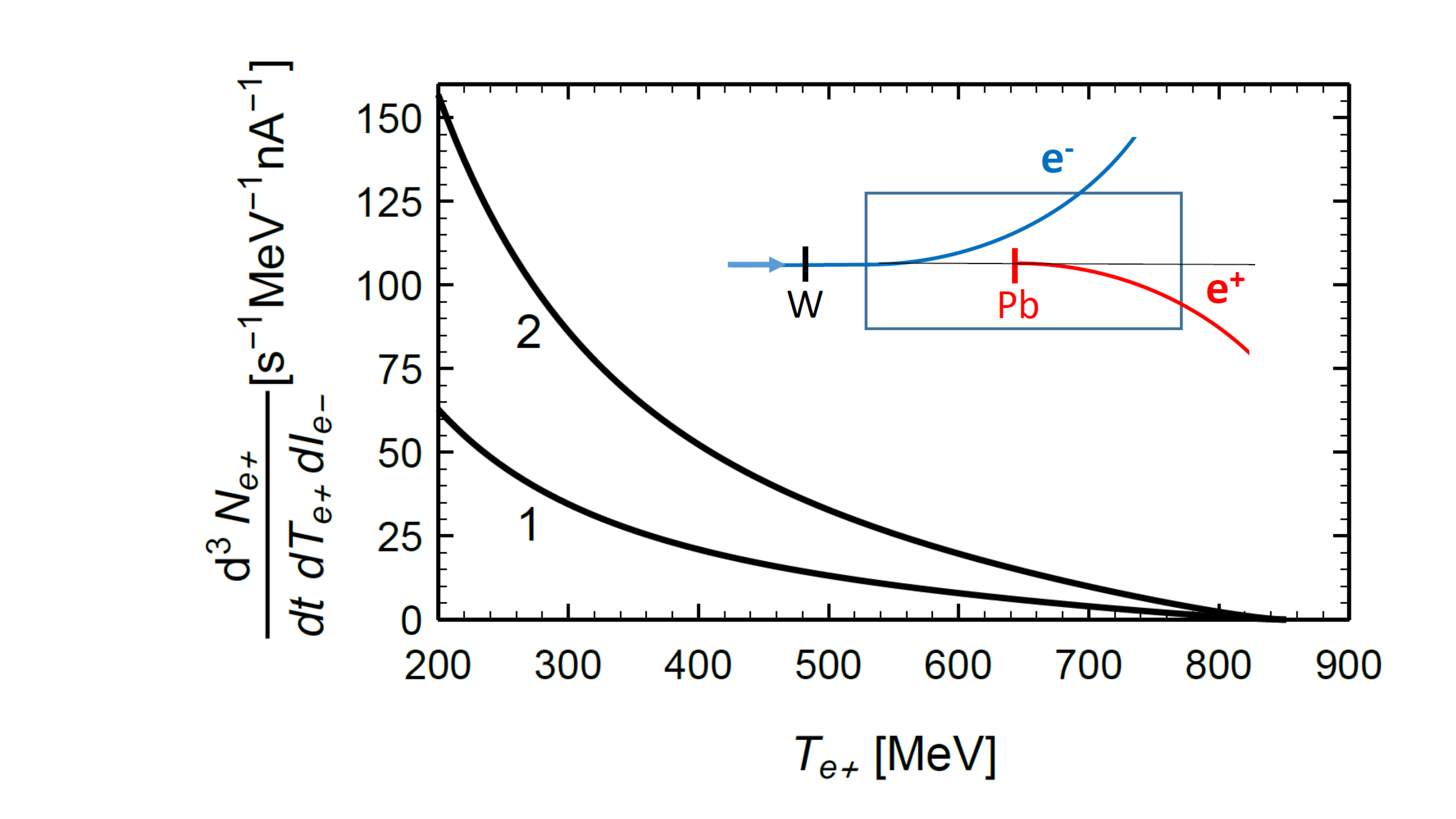}
\caption{
Positron yields: curve 1 for a 10 µm thick tungsten self-converter target, and curve 2 for a 10 µm thick tungsten bremsstrahlung converter and a 20 µm thick lead positron converter target in the separated-converter geometry shown in the inset. Numbers are given per second, 1 MeV kinetic positron energy interval, and an electron beam current of 1 nA.
}
\label{positronYieldWPbInset}  
\end{figure}

The positron beam line will be constructed at the X1 applied physics site of MAMI, see Fig. \ref{Exp-setup}. The primary 855 MeV electron beam hits a 10 $\mu$m thick amorphous tungsten target (Z=74) in which, in a first step, bremsstrahlung photons are produced. In a second step, these bremsstrahlung photons are pair converted. There are two possibilities to produce positrons. In the first one, called in the following the self-converter geometry, the target is at the same time, both, the bremsstrahlung photon source and the positron converter. In the second one, called the separated-converter geometry, the tungsten target is the bremsstrahlung converter located upstream the bending magnet BM1, and the lead positron converter is located inside BM1, see inset of Fig. \ref{positronYieldWPbInset}. The positron converter is shifted into the BM1 by a certain distance in such a manner that the deflected primary electron beam does not scratch the positron converter. In this case study all considerations were done for a kinetic energy $T_{+}= 500$~MeV of the positron beam.

In order to estimate the number of produced positrons, the number of photons created by the electron beam in the bremsstrahlung converter target and the cross-section for positron creation in the converter target must be known.

\begin{figure}[tb]
\centering
\includegraphics*[angle=0,scale=0.5,clip]{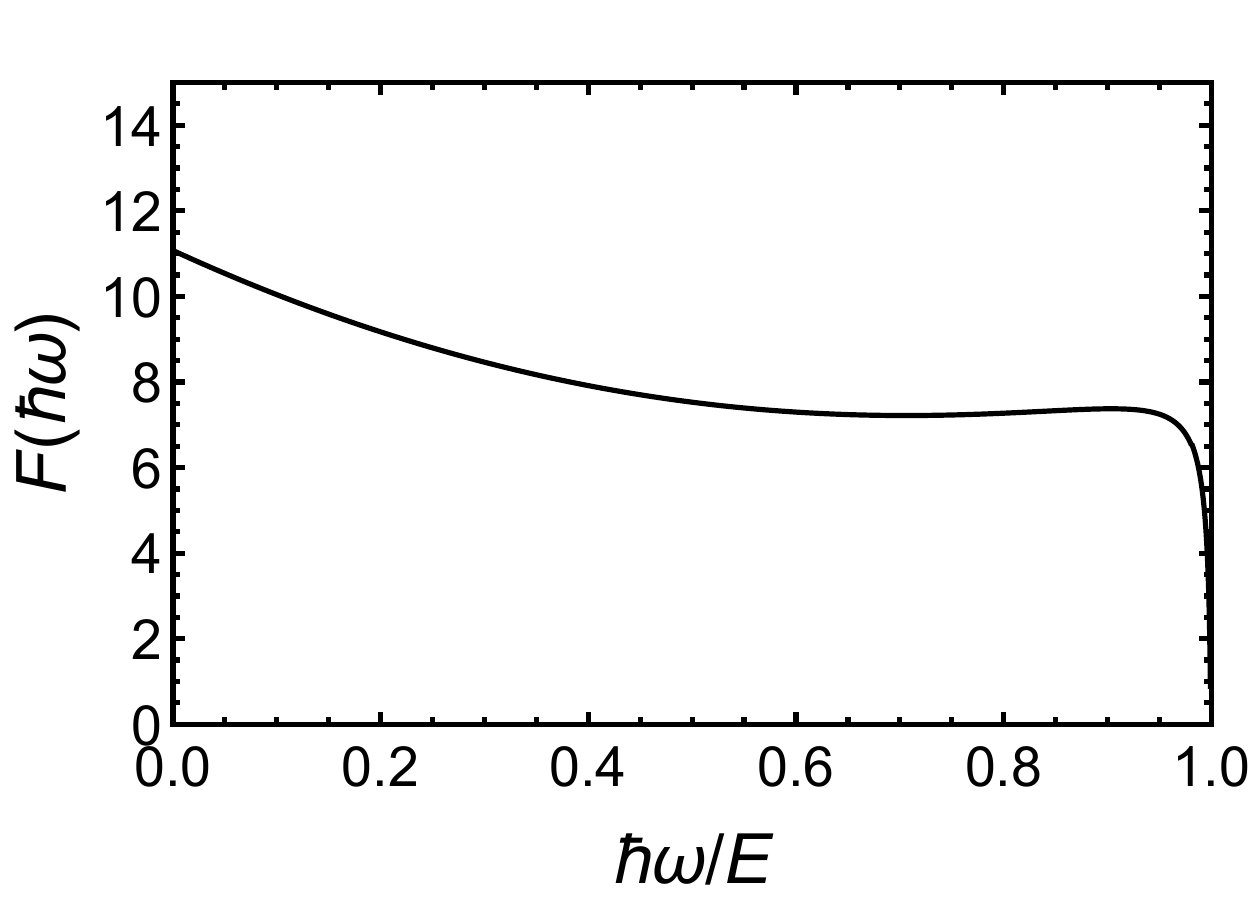}
\caption{
Function $F(\hbar \omega)$ as calculated with Eq. (\ref{EQ__2_}) for electrons with an energy $E$ = 855 MeV impinging on tungsten ($Z$ = 74).
}
\label{FKochMotz}  
\end{figure}

The 3-fold differential bremsstrahlung photon number $d^3 N_0$ created in the target per electron, bremsstrahlung energy interval $d \hbar \omega$, solid angle $d\Omega$, and target density-thickness product $d(\rho t)$ can be written in the case of an ultra-relativistic electron beam with Lorentz factor $\gamma\gg 1$ as
\begin{equation} \label{EQ__1_}
\begin{split}
& \frac{d^{3} N_{0} }{d\hbar \omega \; d\Omega \; d(\rho t)} =
\\
& \frac{N_{A} }{M_{m} } \alpha Z^{2} \; r_{e}^{2} \; \frac{F(\hbar \omega )}{\hbar \omega } \; \Bigg(\frac{3}{2\pi } \gamma ^{2} \frac{1+\gamma ^{4} \theta ^{4} }{(1+\gamma ^{2} \theta ^{2} )^{4} }\Bigg)
\end{split}
\end{equation}
with
\begin{equation} \label{EQ__2_}
\begin{split}
& F(\hbar \omega )=
\\
& 4\left[\left(1+\frac{(E-\hbar \omega )^{2} }{E^{2} } \right)\left(\frac{\Phi _{1} (g)}{4} -\frac{1}{3} \ln \left(Z\right)-f(Z)\right)\right.
\\
& \left.
-\frac{2}{3} \frac{(E-\hbar \omega )}{E} \left(\frac{\Phi _{2} (g)}{4} -\frac{1}{3} \ln \left(Z\right)-f(Z)\right)\right].
\end{split}
\end{equation}
Here are
\[\begin{array}{l}
 {r_{e} =\frac{e^{2} /4\pi \varepsilon _{0} }{m_{e} c^{2} } :{\rm \; classical\; electron\; radius}} \\
 {E:{\rm \; total\; energy\; of\; electron\; beam}} \\
 {N_{A} :{\rm Avogadro\; constant}} \\
 {M_{M} :{\rm \; molar\; mass\; of\; the\; target}} \\
 {Z:{\rm \; atomic\; number\; of\; the\; target}}
 \end{array}\]
Eqs. \eqref{EQ__1_} and \eqref{EQ__2_} have been taken from Koch et al., \cite[Formula 3CS]{Koch:95}. In Eq. \eqref{EQ__1_} the last term in big parentheses is the angular dependence term which has been included. The screening functions $\Phi _{1} (g)$, $\Phi _{2} (g)$ were taken from Motz et al. \cite[Formula 3D -- 1003]{Motz:69} with $g = 100\cdot\hbar \omega\cdot m_ec^2/E (E - \hbar \omega ) Z^{1/3}$, and  $f(Z)$ from \cite[Table 6.05]{Motz:69}.
The solid angle integrated Eq. \eqref{EQ__1_} reads
\begin{equation} \label{EQ__3_}
\frac{d^{2} N_{0} }{\hbar \omega \; d(\rho t)} =\frac{N_{A} }{M_{m} } \alpha Z^{2} \; r_{e}^{2} \; \frac{F(\hbar \omega )}{\hbar \omega } \; .
\end{equation}
The result has been checked with the tables of Seltzer and Berger \cite{Seltzer:86} for lead (Z = 82) at a beam energy of 1~GeV. An agreement of better than a few percent was found. The function $F(\hbar \omega)$ is shown in Fig. \ref{FKochMotz}.

\begin{figure}[b]
\centering
\includegraphics*[angle=0,scale=0.5,clip]{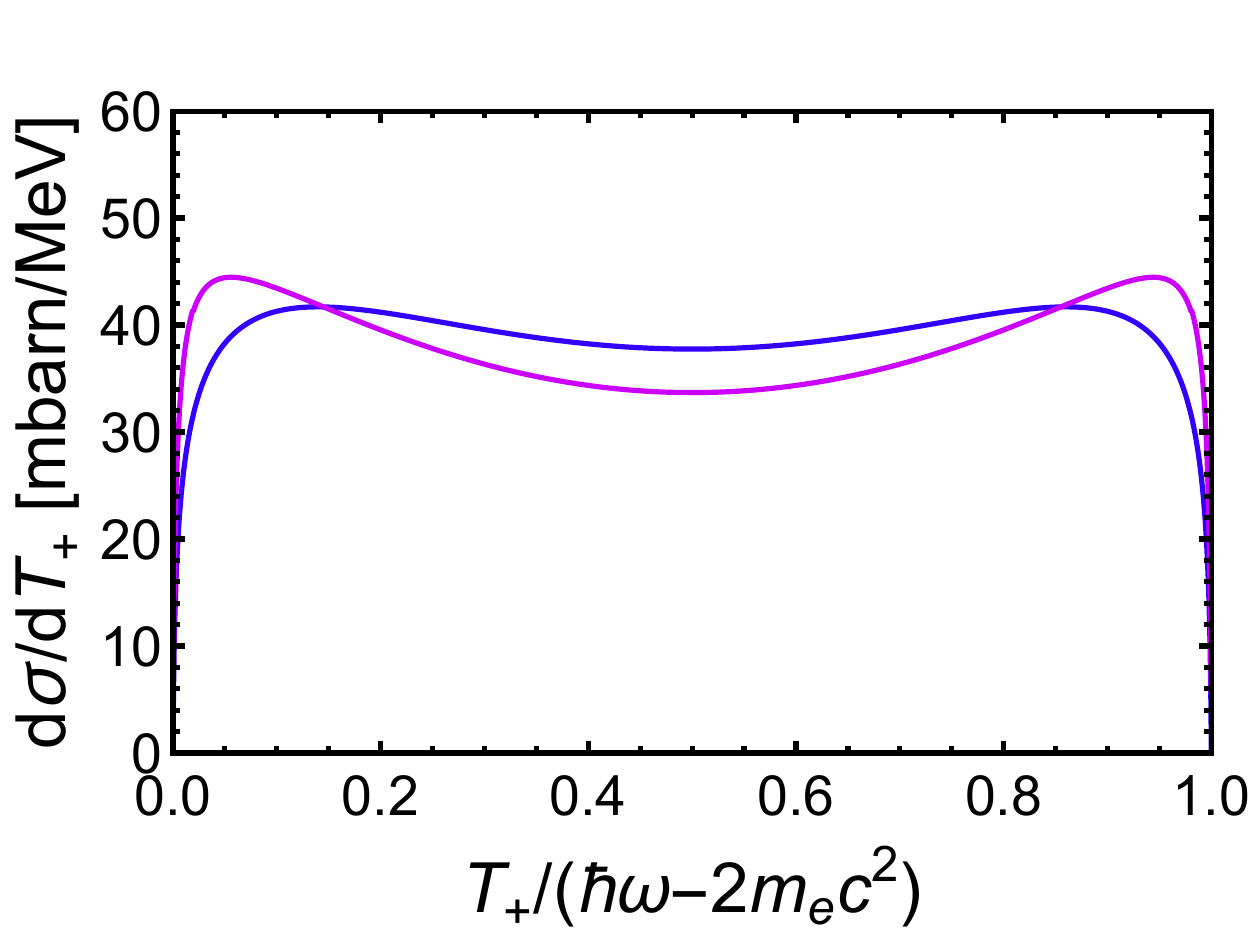}
\caption{
Differential positron production cross section at a photon energy of 855 MeV. The violet curve is from Motz et al. \cite[Formula 3D -- 1009]{Motz:69}, the blue one from Hubbell et al. \cite[ formula (4)]{Hubbell:80}.
}
\label{differential_cross_section}  
\end{figure}

For the calculation of the positrons created in the converter target two approaches have been considered: (i) the Davis-Bethe-Maximon formula of Motz et al. \cite[Formula 3D -- 1009]{Motz:69}, and (ii) the Davies, Bethe, and Maximon formula for the unscreened differential cross-section according to Hubbell et al. \cite[Eq. (4)]{Hubbell:80}, multiplied by a screening correction factor \textit{f${}_{sc}$}:
\begin{eqnarray} \label{crossSectionPositrons}
\frac{d\sigma}{dT_{+}}(T_{+},\hbar \omega) =
\nonumber\\
& &\hspace{-1.8 cm}
\alpha Z^2 r_e^2 \frac{1}{(\hbar \omega)^3}\Big((T_{+}+m_e c^2)^2+(T_{-}+m_e c^2)^2
\nonumber\\
& &\hspace{-1.8 cm}
+\frac{2}{3}(T_{+}+m_e c^2)(T_{-}+m_e c^2)\Big)
\nonumber\\
& &\hspace{-1.8 cm}
\times\Big(2\ln\frac{2(T_{+}+m_e c^2)(T_{-}+m_e c^2)}{ m_e c^2\hbar \omega}
\nonumber\\
& &\hspace{-1.8 cm}
-1-2f(Z)\Big)f_{sc}(\hbar \omega).
\end{eqnarray}
The screening correction factor
\begin{equation} \label{fsc}
f_{sc}(\hbar \omega)=\sigma_{pair}(\hbar \omega)\Big/\int\limits_{T_{+}=0 }^{\hbar \omega-2m_{e} c^{2}} \frac{d\sigma}{dT_{+}}(T_{+},\hbar \omega)dT_{+}.
\end{equation}
is the total cross-section $\sigma_{pair}(\hbar \omega)$ in the nuclear field as taken from the NIST tables \cite{BerH:NIST} and the integral of the differential cross-section Eq. (\ref{crossSectionPositrons}) over all positron energies. Both approaches differ somehow in the shape, see Fig. \ref{differential_cross_section}, and also in the total cross sections by about 10-20 \%. It is not clear which approach is the better one. Somehow arbitrarily, all calculations have been performed with Eq.  (\ref{crossSectionPositrons}).

The positron yield has been calculated for various kinetic positron energies \textit{T${}_{+}$} by the integral
\begin{equation} \label{eq__8_}
\begin{split}
  & \frac{dN_{e+} }{dT_{+}} (T_{+})= \frac{N_{A} }{M_{m} } \rho _{C} t_{C}\times
 \\
 &  \int\limits_{\hbar \omega =T_{+} +2m_{e} c^{2} }^{855\; MeV}\frac{d\sigma }{dT_{+} } (T_{+},\hbar \omega )\; \frac{dN_{0} }{d\hbar \omega } (\hbar \omega )\;  d\hbar \omega
 \end{split}
  \end{equation}
The converter target with density \textit{$\rho$${}_{C}$} and thickness \textit{t${}_{C}$} may either be a separated-converter target of, e.g., lead, or the tungsten target with thickness \textit{t${}_{W}$} itself. In the latter case, the converter target thickness is $t_C = t_{W}$/2.

Calculated yields are shown in Fig. \ref{positronYieldWPbInset}. Notice, the lower the positron energy, the higher the yield. The reason is the increasing number of bremsstrahlung photons above the threshold $T_{+} + 2 m_ec^{2}$.

\section{Positron transfer matrix of bending magnet BM1} \label{ImagingBM1}

The only non standard element in the positron beam line to be constructed is the bending magnet BM1. In this section ray trace calculations will be described to obtain its transfer matrix.

The bending magnet BM1 is a modified version of a $\theta_0$ = 30$\mathrm{{}^\circ}$ magnet with a bending radius $R$ = 3820 mm originally used at the electron-positron storage ring DCI in Saclay. The iron return yoke was placed inside the electron orbit, meaning that positrons can leave the magnet into the free space outside. The gap was reduced to 60 mm to reach a magnetic field of $B$ = 1.0517 Tesla which is required to achieve the bending angle of 43.53° for guiding the electrons into the beam dump.

The natural entrance point of the magnet has a distance from its nominal orbit symmetry point, in the middle of the magnet and the middle of the pole piece, of $R \sin(\theta_0/2)$ and $R(1-\cos(\theta_0/2))$ in longitudinal and transverse direction, respectively. The magnet is rotated counter clockwise by 21.765°. A cartesian coordinate system $(x_{L}, y_{L},z_{L})$ has been chosen with the $z_L$ axis parallel to the beam direction, the $x_L$ axis in the plane of the electron orbit pointing into the direction as shown by the coordinate system in Fig. \ref{Exp-setup}. The origin of the coordinate system is displaced from the natural entrance point by (22.8, 0, -2.7) mm.

\begin{figure}[]
\centering
\includegraphics*[width=200pt]{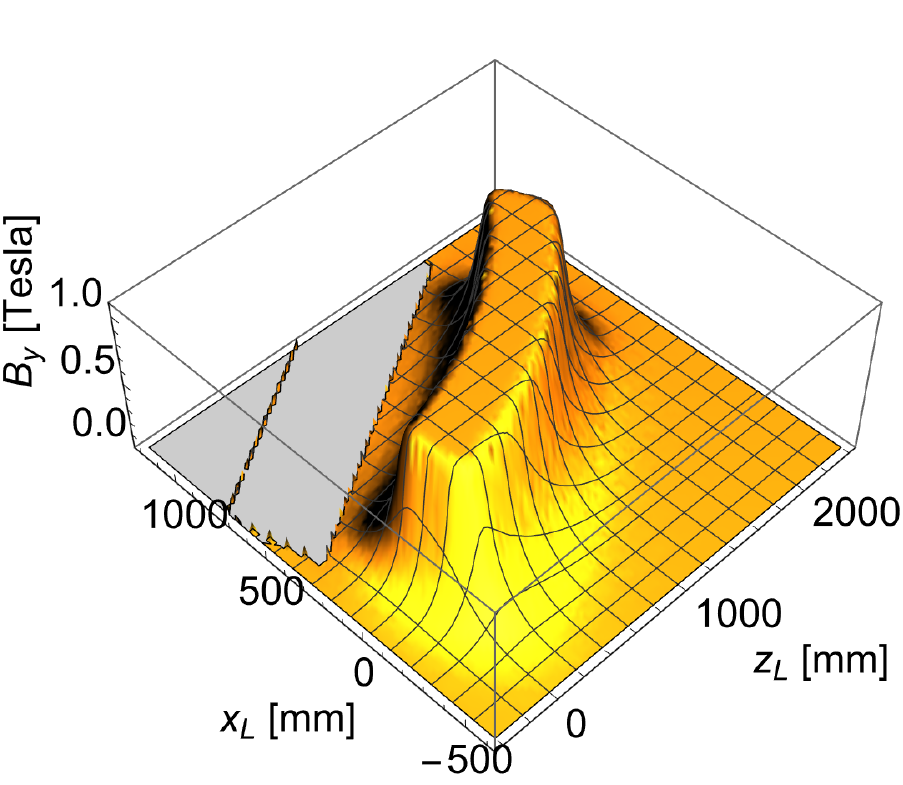}
\caption{
Magnetic field component $B_{y} (x_L, z_L)$ of the BM1 magnet in the symmetry plane $(y_L = 0)$  as function of the  horizontal $x_{L}$ and $z_{L}$ coordinates in the laboratory system. The maximum magnetic field is $B$ = 1.0517 Tesla. The front face of the magnet is tilted counter clockwise by an angle of 6.765°. The electron beam enters the magnet at $x_{L}$ = 0 and $z_{L}$ = 0. The gray area indicates the negative magnetic field in the iron return yoke.}
\label{magnetic_field}   
\end{figure}

To obtain nominal orbits, ray trace calculations were performed by solving the coupled differential equations in the laboratory cartesian coordinate system $(x_{L}, y_{L},z_{L})$
\begin{equation} \label{EQ__9_}
\begin{split} z_{L} ''(s)-\frac{q c}{mc^{2} } B_{y}\big(x_{L} (s),z_{L}(s)\big) x_{L} '(s)=0
\\\\
x_{L} ''(s)+\frac{q c}{mc^{2} } B_{y} \big(x_{L} (s),z_{L} (s)\big)z_{L} '(s)=0.
\end{split}
\end{equation}
The vertical magnetic field component ${B}_{y} (x_L, z_L)$ is shown in Fig. \ref{magnetic_field} in the horizontal symmetry $(x_{L}, 0, z_{L})$ plane. It is directed vertically into the direction of the $y_{L} \equiv y$ coordinate. The magnetic field was calculated, based on the geometrical and material configuration of the magnet, with the code 'CST - studio suite' \cite{CST-studio:22}.
The quantity \textit{s} = \textit{v t} $\approx$ c \textit{t} is the path length of the trajectory (\textit{x${}_{L}$}(s), \textit{z${}_{L}$}(\textit{s})) which is the solution of Eqs. \eqref{EQ__9_} in the horizontal plane, i.e. for \textit{y${}_{L}$}(\textit{s}) = 0. The charge $q$ is -$e$ for electrons and +$e$ for positrons, with $e$ the elementary charge. The quantity \textit{m}c${}^{2}$ = \textit{E} is the total energy of the electrons (positrons).
The correctness of the solution of Eqs. \eqref{EQ__9_}, which have been performed with the Mathematica 10.4 package, was checked by calculating the trajectory of the 855 MeV electron beam. A deflection angle of 43.70° resulted which is rather close to the nominal angle of 43.53°.

The W target is positioned in the self-converter geometry inside the magnet at $x_{L,W}$ = 12.37 mm, $z_{L,W}$ = 196.0 mm. The lateral displacement is due to the curvature of the electron orbit which makes at the target position an angle of 5.205° with respect of the $z_L$ axis. The nominal 500 MeV positron beam exits the BM1 vacuum chamber, see Fig. \ref{Exp-setup}, at  $z^{(vc)}_{L}$ = 1682.5 mm, which is the distance from the entrance of the BM1 magnet and the flange of the vacuum chamber, at $x^{(vc)}_{L}$= -349.6 mm.

\begin{figure}[tb]
\centering
\includegraphics*[width=170pt]{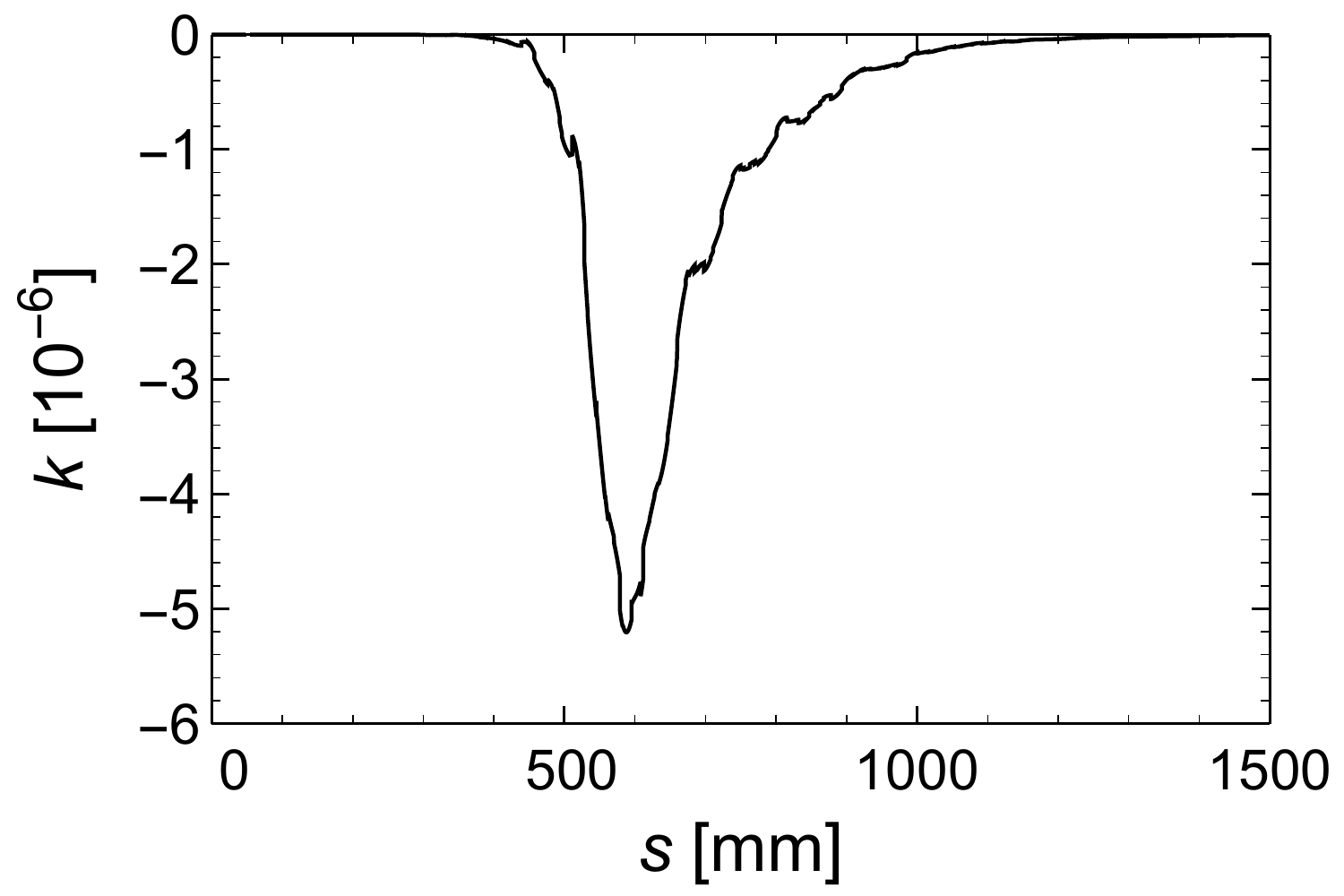}
\caption{
Quadrupole strength $k(s)$ of the bending magnet BM1 as function of the path length of the trajectory $s$ for a positron energy of 500 MeV.The fluctuation are due to numerical inaccuracies in the calculation of the magnetic field.
}
\label{quadrupoleStrengthBM1}
\end{figure}

Deviations from the nominal trajectory were calculated by solving the differential equations
\begin{equation} \label{EQ__10_}
\begin{split}
 & x''(s)+\left(\frac{e^{2} }{p^{2} } B_{y}^{2} \left(\vec{r}(s)\right)-k(s)\right)x(s)=0
\\
 & y''(s)+k(s)y(s)=0
\end{split}
\end{equation}
with
\begin{equation}
\begin{split}
 k (s)   =  &\frac{e}{p} \hat{n}_{L} (s)\frac{\partial B_{y } (\vec{r}_{L} )}{\partial \vec{r}_{L} }
\\
  =  &\frac{e}{p} \left(-t_{L,z} (s)\frac{\partial B_{y } \left(x_{L} (s),y_{L} (s)\right)}{\partial x} + \right.
 \\
 & \left.  t_{L,x} (s)\frac{\partial B_{y } \left(x_{L} (s),y_{L} (s)\right)}{\partial z} t_{x} \right)
\end{split}
\end{equation}
and $\vec{r}(s)=\left(x(s),y(s)\right)$ the deviation from the nominal trajectory counted perpendicular to the trajectory \cite{Wille:92}. To obtain the quadrupole strength \textit{k${}$}(\textit{s}), depicted in Fig. \ref{quadrupoleStrengthBM1}, the directional derivative perpendicular to the nominal trajectory must be known. It was calculated in the laboratory frame with the tangential unit vector $\hat{t}_{L} (s)=\left(t_{L,x},\; t_{L,z}\right)=$ $\left(dx_{L}/ds,dz_{L}/ds\right)$ after a rotation in the horizontal plane by an angle of -- 90$\mathrm{{}^\circ}$ yielding $\hat{n}_{L}=(-t_{L,z},t_{L,x})$.  With $\partial B_{y } (\vec{r}_{L} )/\partial \vec{r}_{L} $the gradient of the scalar field $B_{y } (\vec{r} )$ in the laboratory system is denoted.

The transfer matrix for the column vector $(x,x',y,y',\Delta p/p)$ reads in general form with $mh_{ik}$ and $mv_{ik}$ the horizontal and vertical matrix elements, respectively,

\begin{equation}
M= \left(
\begin{array}{ccccc}
 mh_{11} & mh_{12} & 0 & 0 & mh_{13} \\
 mh_{21} & mh_{22} & 0 & 0 & mh_{23} \\
 0 & 0 & mv_{11} & mv_{12} & 0 \\
 0 & 0 & mv_{21} & mv_{22} & 0 \\
 0 & 0 & 0 & 0 & 1 \\
\end{array}
\right)
\end{equation}
or numerically for 500 MeV positrons exiting BM1 at $z^{(vc)}_{L}$ = 1682.5 mm
\\
$
M_{BM1}=
\\\\
\small
 \left(
\begin{array}{ccccc}
 -0.002845 & 975.931 & 0 & 0 & 397.63 \\
 -0.001026 & 0.3691 & 0 & 0 & 0.2992 \\
 0 & 0 & 1.8536 & 2086.76 & 0 \\
 0 & 0 & 0.0009777 & 1.6399 & 0 \\
 0 & 0 & 0 & 0 & 1 \\
\end{array}
\right).
$
\\\\
The (2$\times$2) submatrices have the structure of a quadrupole. Since the element $k(s)$ is negative, see Fig. \ref{quadrupoleStrengthBM1}, consequently also the element $mh_{21}$ turns out to be negative, and the bending magnet behaves for positrons like a horizontally focusing quadrupole, and at the same time, vertically as de-focusing one. The dispersion is for 500 MeV  $pv_{+}/mh_{13}$ = 1.257 MeV/mm.

\section{Experimental tests of calculations}
\label{experimentaltests}
\begin{figure}[b]
\centering
\includegraphics*[width=200pt]{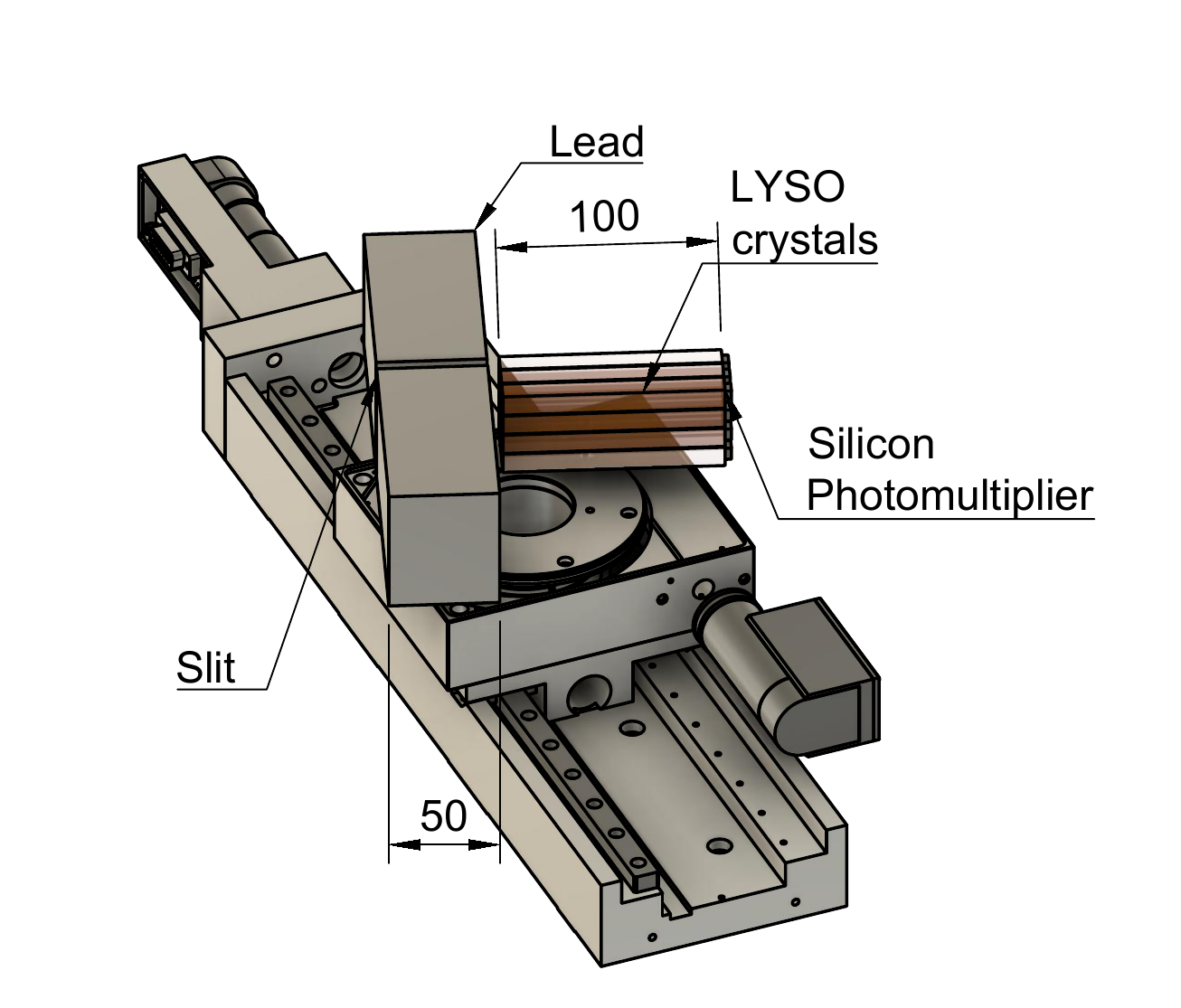}
\caption{
Detector assembly for the measurements of positrons. A number of 12 LYSO crystals serve as a calorimeter. Each of the crystals has a size of 100x10x10~mmm$^3$, read out with 6 x 6 mm silicon photomultipliers operating in the Geiger mode. The radiation length of LYSO is 11.4~mm and the Moliere radius 20.7~mm. Data acquisition is performed with a fast 125~MHz sampling ADC. The aperture in front of the detector is made of 50 mm thick lead with a 6 mm wide slit. The support of the detector assembly allows rotation, and linear movement along the dispersive plane.
}
\label{lyso}
\end{figure}

Since the calculation of the transfer matrix for positrons for the magnet BM1, as described in the last section, turned out to be rather complex, it appeared to us mandatory to test the results. An experiment was performed behind the vacuum chamber of the bending magnet BM1, see Fig. \ref{Exp-setup}, with which not only the imaging properties of BM1 could be checked but also the calculation of the positron yield described in section \ref{Positron_creation}.

\begin{figure}[]
\centering
\includegraphics*[width=200pt]{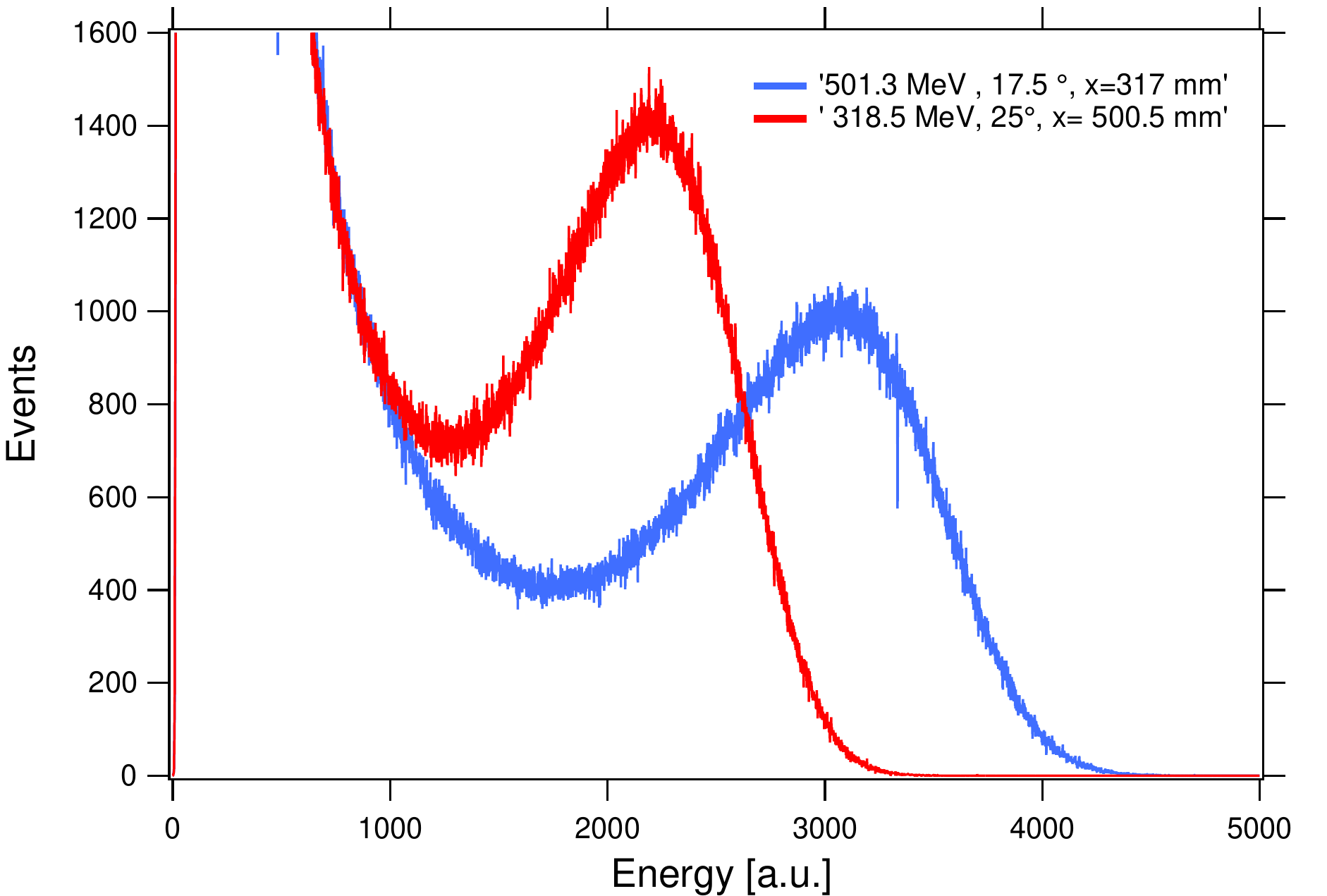}
\caption{
Energy spectra for positrons taken with the detector assembly of Fig. \ref{lyso} at  $z^{(vc)}_{L}$ = 1682.5 mm for $x_D$ = 317~mm (blue line) and 500.5~mm (red line). Tungsten target thickness 10~µm, electron beam current 14.2 nA, data collection time 1,000 s. According to an energy calibration with electrons of the MAMI accelerator with energies of 855 and 350 MeV, the peak energies are 318.5 and 501.3 MeV.
}
\label{exp-result} 
\end{figure}
\begin{figure}[b]
\centering
\includegraphics*[width=200pt]{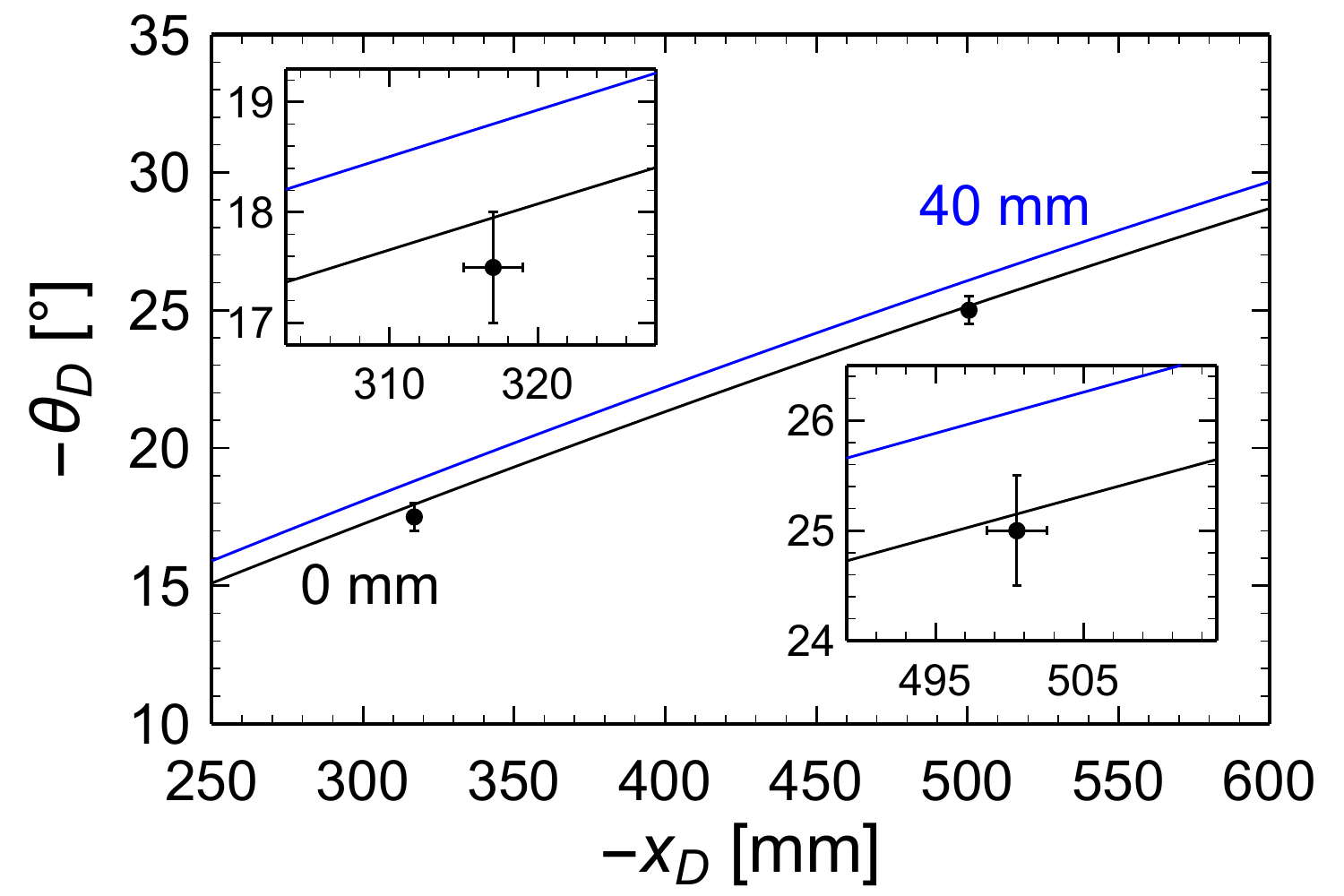}
\caption{
Angle and position measurements with the setup shown in Fig. \ref{lyso} at $z^{(vc)}_{L}$ = 1682.5 mm. Values are $-x_D$ = (317 $\pm$ 1)~mm and (500.5 $\pm$ 1)~mm with angles of $-\theta_D$ = (17.5 $\pm$ 0.5)° and (25.0 $\pm$ 0.5)°, respectively. The black line "0 mm" indicates calculations at an optimized tungsten-target position, the blue one "40 mm" with a 40 mm downstream shifted target. A detailed comparison is depicted in the insets.
}
\label{AnglePositionMesurements}
\end{figure}

Positrons from a 10~$\mu$m thick W target, located in the BM1 as described in section \ref{ImagingBM1}, were analyzed with a detector assembly shown in Fig. \ref{lyso}. Spectra taken with the LYSO calorimeter are shown in Fig.  \ref{exp-result}. A special feature of the detector assembly is the fact that both, the lateral position $x_D$ of the positron beam at the position of the flange of the vacuum chamber at $z_D$ = 1,682.5 mm, and via a rotation also its angle $\theta_D$ with respect to the $z_L$ axis can be measured. The results for two measurements are shown in Fig. \ref{AnglePositionMesurements}. The tungsten target position was adapted in the calculation to match the measurement. This way also the energy of the positron beam could be determined since the whole construction is a magnetic spectrometer.

\begin{figure}[h]
\centering
\includegraphics*[width=220pt]{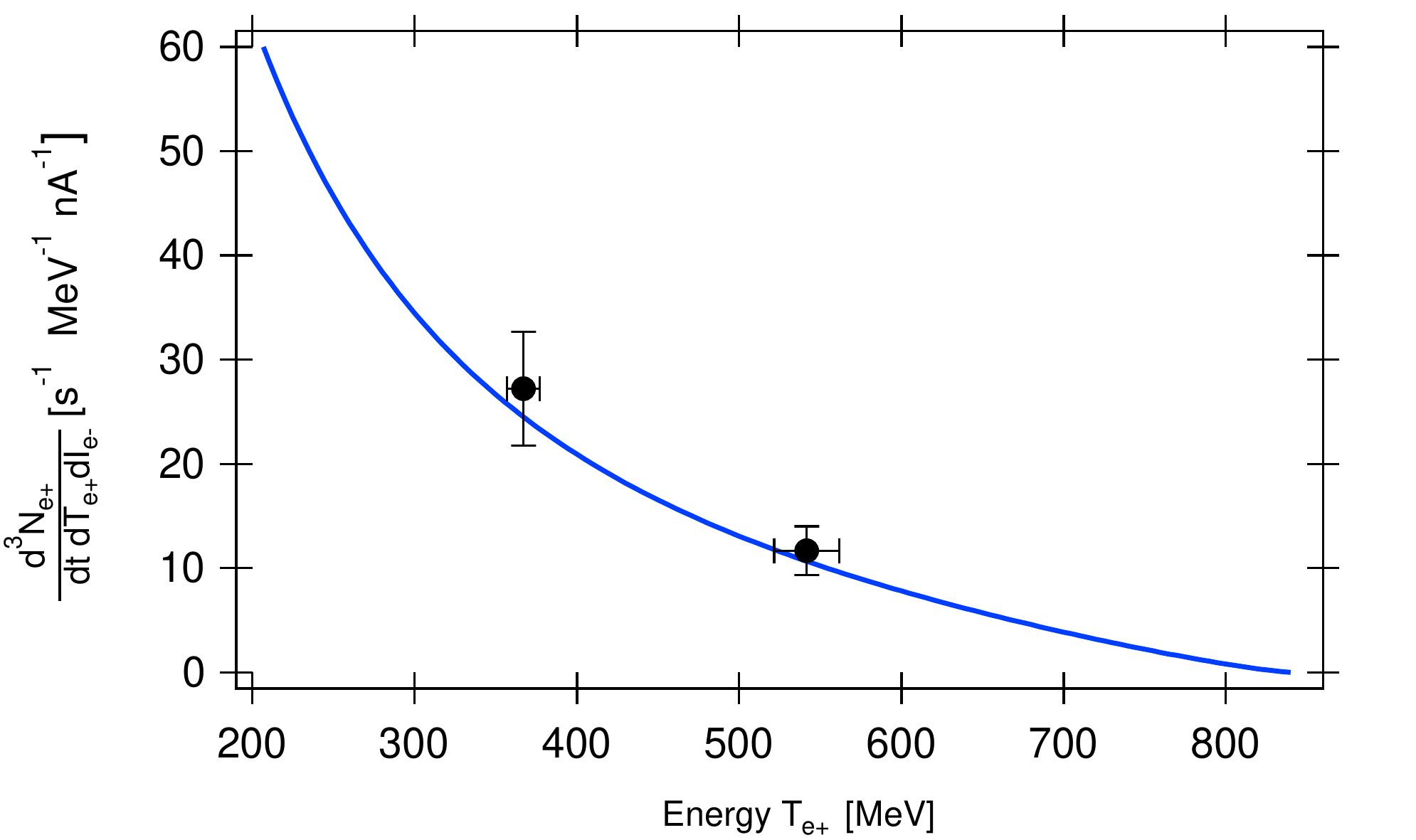}
\caption{
Calculated positron yield for a 10 µm thick tungsten self-converter target, blue line, as function of the positron energy $T_+$. The error bars are measurements with the detector shown in Fig. \ref{lyso} at the spectrometrically determined energies of (367.2 $\pm$ 8.9) MeV and (541.6 $\pm$ 16.5) MeV, and dispersions of 0.7268 MeV/mm and 1.439 MeV/mm, respectively.
}
\label{positronYieldExp}  
\end{figure}

Measurements of the positron yields for two spectrometrically determined energies are shown in Fig. \ref{positronYieldExp} which agree well with the calculations described in section \ref{Positron_creation}.

The experimental test of the previously described calculations was of particular importance since the beam line to be constructed requires substantial funds.

\section{Ray trace calculations} \label{raytracing}
\subsection{Self-converter geometry} \label{self-converting-section}

The simplest way to produce a positron beam is to position the target in the electron trajectory within BM1 as described in the previous section \ref{ImagingBM1}. Bremsstrahlung photons created in the target are pair converted in the target itself. The strongly forward directed positrons will be deflected in the bending magnet BM1 into opposite direction like the 855 MeV electron beam and energy separated. A slit in the horizontal plane cuts out a certain energy band from the continuous energy spectrum, characterized by $\Delta T_{+}$, or $\Delta p/p \approx \pm 1/2 \cdot \Delta T_{+}/ T_{+}$, which will be deflected back by a second bending magnet BM2 and this way guided into the X1 experimental area, see Fig. \ref{Exp-setup}. To focus the positron beam onto a channeling crystal in the single crystal chamber SCCh, a quadrupole doublet has been placed  between the bending magnets BM1 and BM2, i.e., a horizontally defocusing quadrupole Q${}_{D}$(\textit{k,s}) with \textit{k${}_{Q1}$} $\mathrm{>}$ 0 followed by a horizontally focusing one Q${}_{F}$(\textit{k,s}) with \textit{k${}_{Q2}$} $\mathrm{<}$ 0.

\begin{figure}[]
\centering
\includegraphics*[width=100pt]{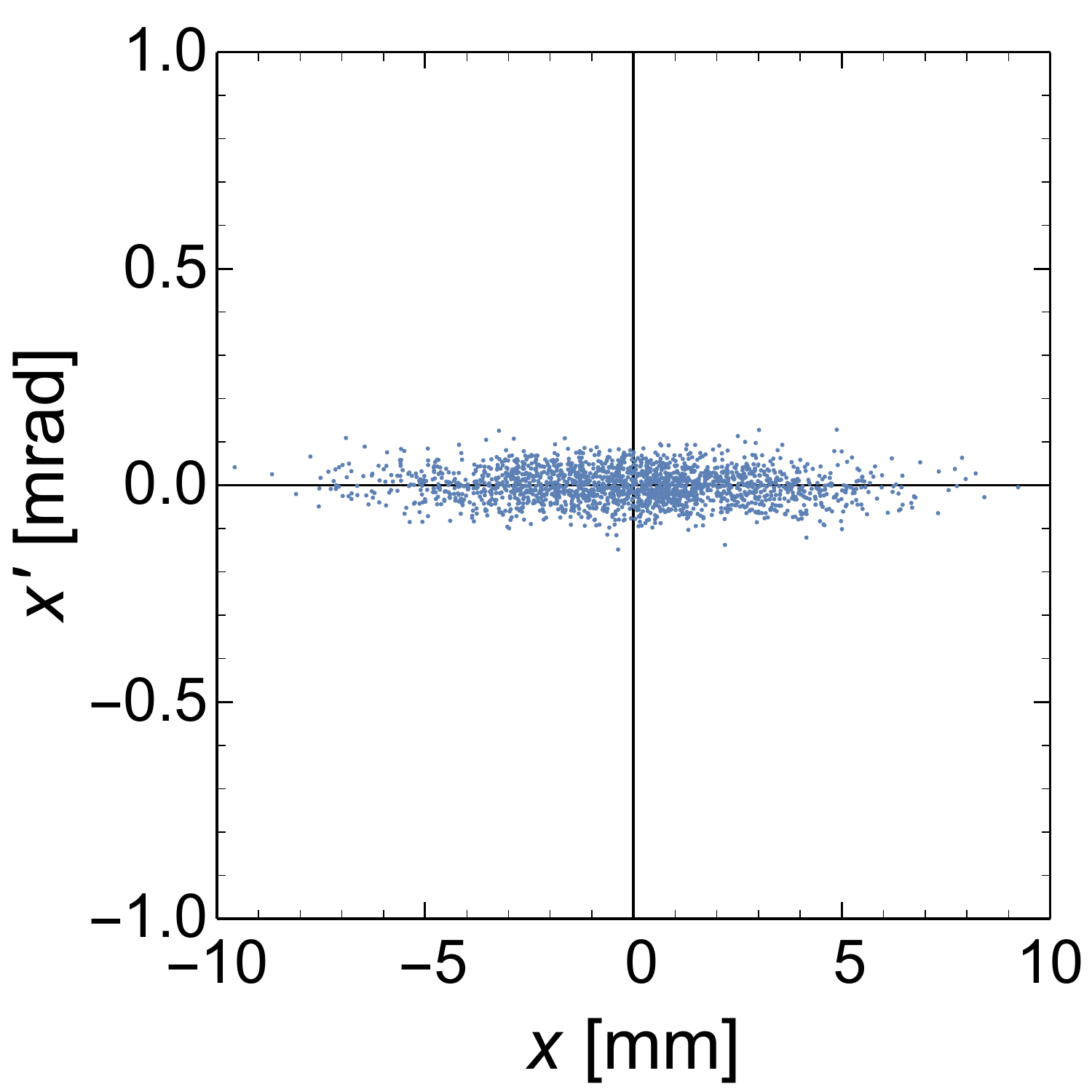} \includegraphics*[width=100pt]{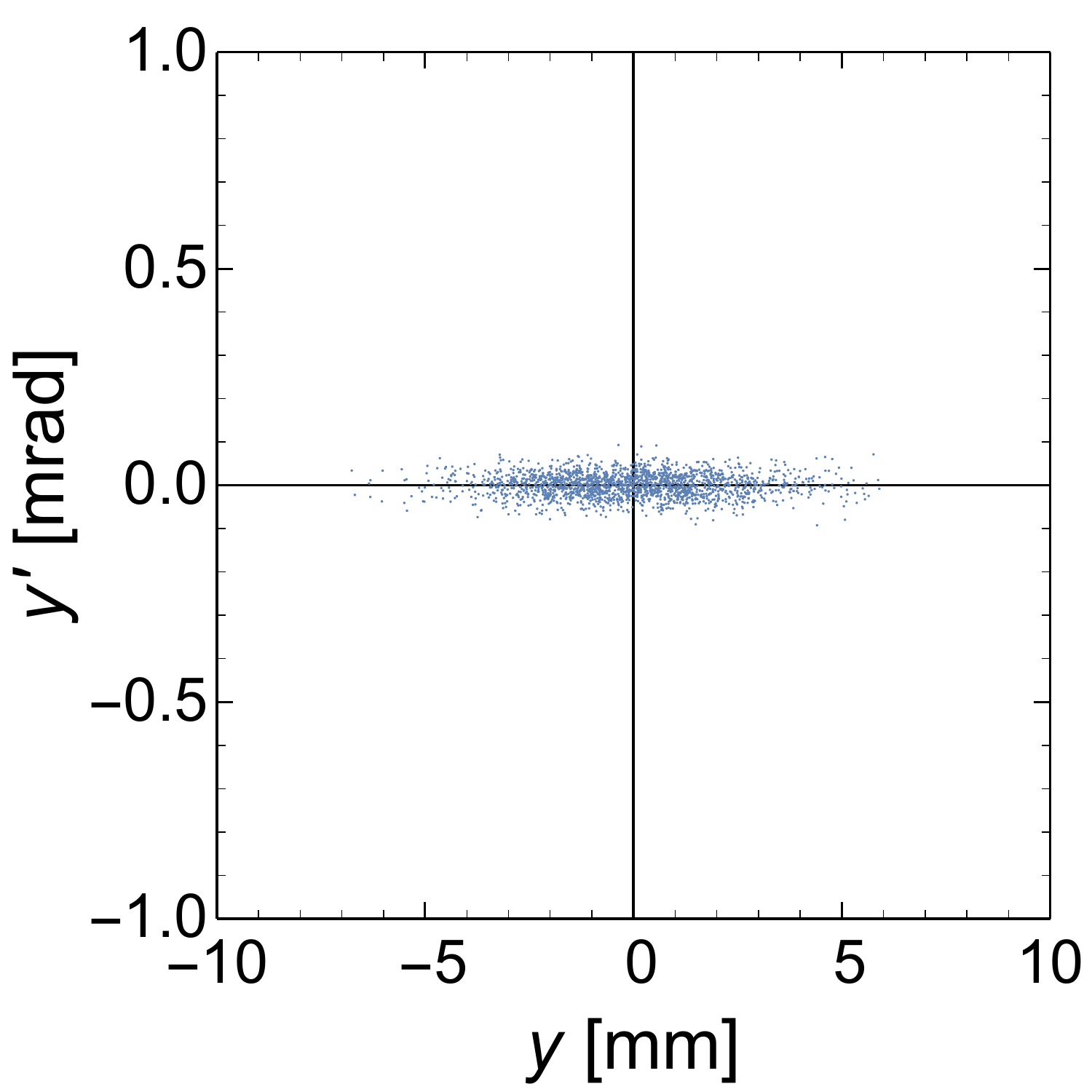}
\caption{
Phase space distributions at the SCCh for $\Delta p/p = 0$, left horizontal, and right vertical. Calculations were performed for 2000 rays randomly distributed at the 10 $\mu$m tungsten target according to Gaussians with standard deviations as described in the Appendix.}
\label{phase_space}  
\end{figure}

\begin{figure}[]
\centering
\includegraphics*[width=100pt]{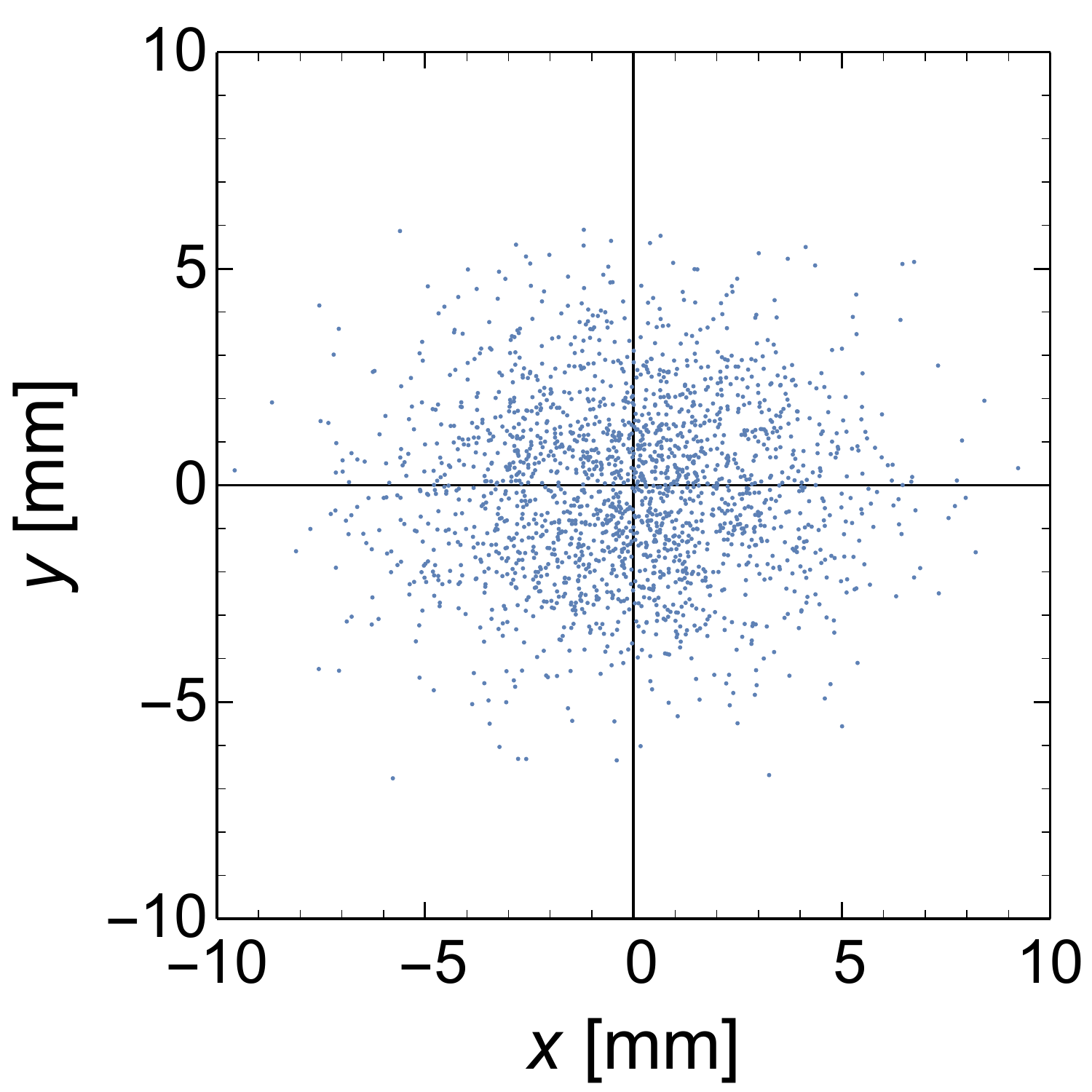}\includegraphics*[width=100pt]{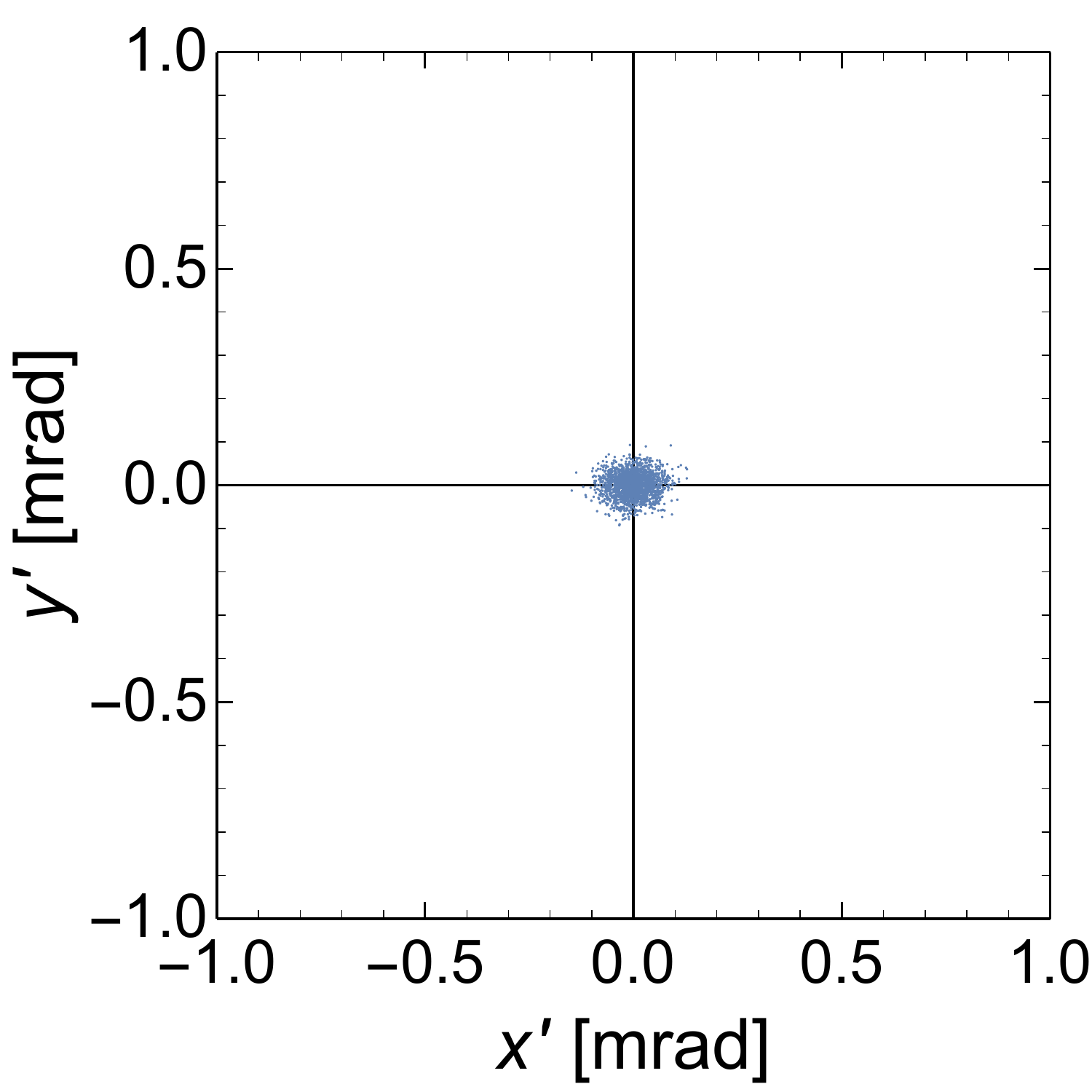}
\caption{
Spot size, left, and angular distribution, right, at the SCCh position for $\Delta p/p = 0$. Projected FWHM are $\Delta x$ = 6.7 mm, $\Delta y$ = 4.9 mm, $\Delta x'$ = 0.094 mrad, $\Delta y'$ = 0.063 mrad.
}
\label{spot_size}  
\end{figure}

With the standard expressions for the drift matrix \textit{S}(\textit{s}) with \textit{s} the drift length, for a defocusing and focusing quadrupole, and for a bending magnet BM2 with length $s_{BM2} = R_{BM2} \cdot\varphi_{BM2}$, $R_{BM2}$ is the bending radius and \textit{$\varphi_{BM2}$} the deflection angle, the total transfer matrix reads

\begin{equation} \label{EQ__13_}
\begin{split}
M_{tot}=
& S(s_{4} )\cdot M_{BM2}^{PE}(s_{BM2},R_{BM2},\psi_{PE})\cdot S(s_{3} )\cdot
\\
& Q_{F} (k_{Q2} ,s_{Q2} )\; \cdot S(s_{2} ) \cdot Q_{D} (k_{Q1} ,s_{Q1} )\cdot
\\
& S(s_{1} ) \cdot M_{BM1}
\end{split}
\end{equation}
In $M_{BM2}^{PE}(s_{BM2},R_{BM2},\psi_{PE})$ the pole edge focusing matrix $M^{PE}(\psi_{PE},R_{BM2})$ is included. With the parameters of the beam optics elements compiled in the Appendix, the phase space ellipses can be aligned at the SCCh for a momentum spread $\Delta p/p=0$ parallel to the horizontal $x$ and simultaneously to the vertical $y$ coordinates. This way the angular spreads $x'$ and $y'$ are minimized. Beam characteristics were calculated with a trial-and- error method with input values for the spacial and angular distributions at the target position which are also compiled in the Appendix. To illustrate the beam quality, phase space distributions $(x,x')$ and $(y,y')$  at the SCCh are shown in Fig. \ref{phase_space}. In Fig. \ref{spot_size} the corresponding spot size and angular distribution are depicted.

\begin{figure}[]
\centering
\includegraphics*[width=100pt]{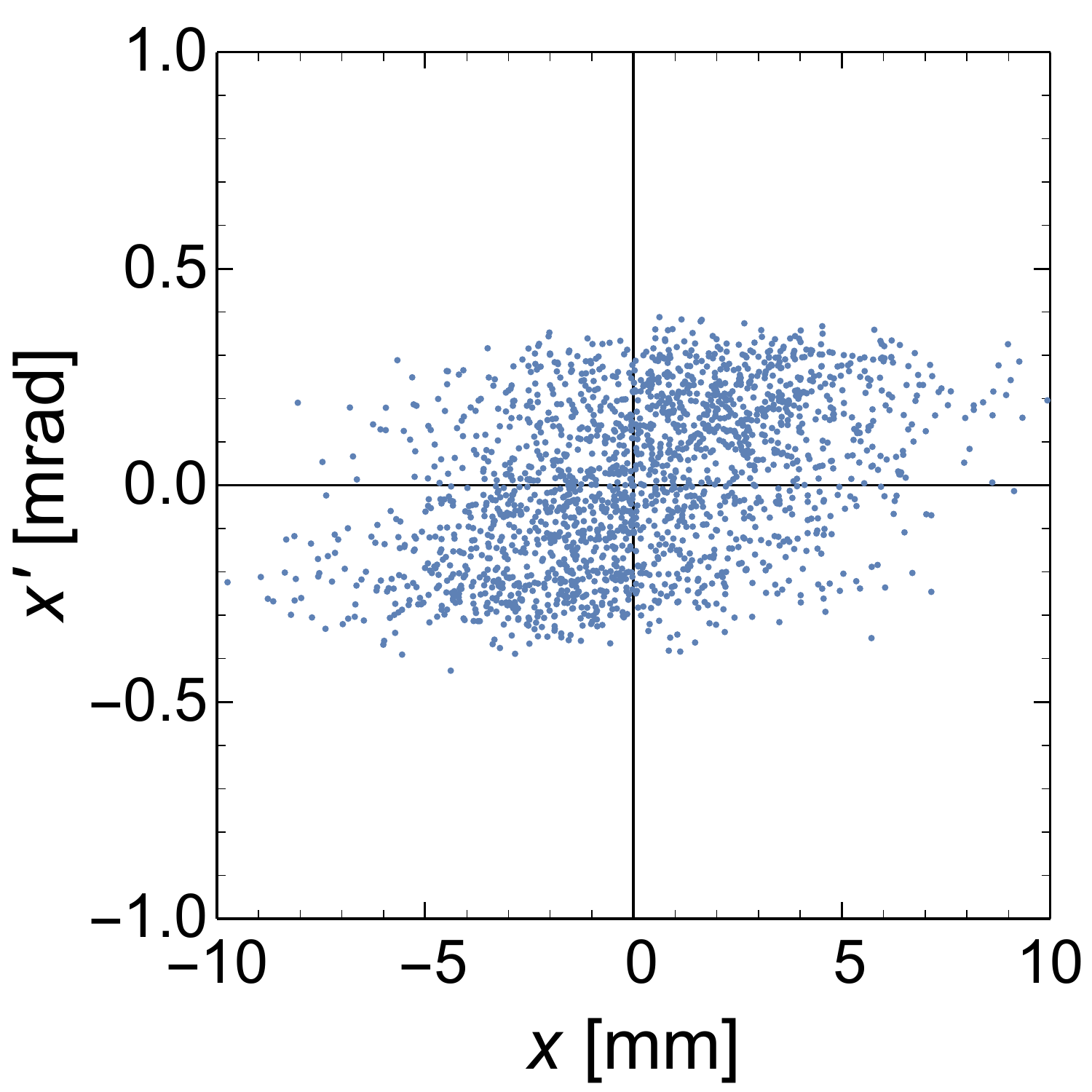}\includegraphics*[width=100pt]{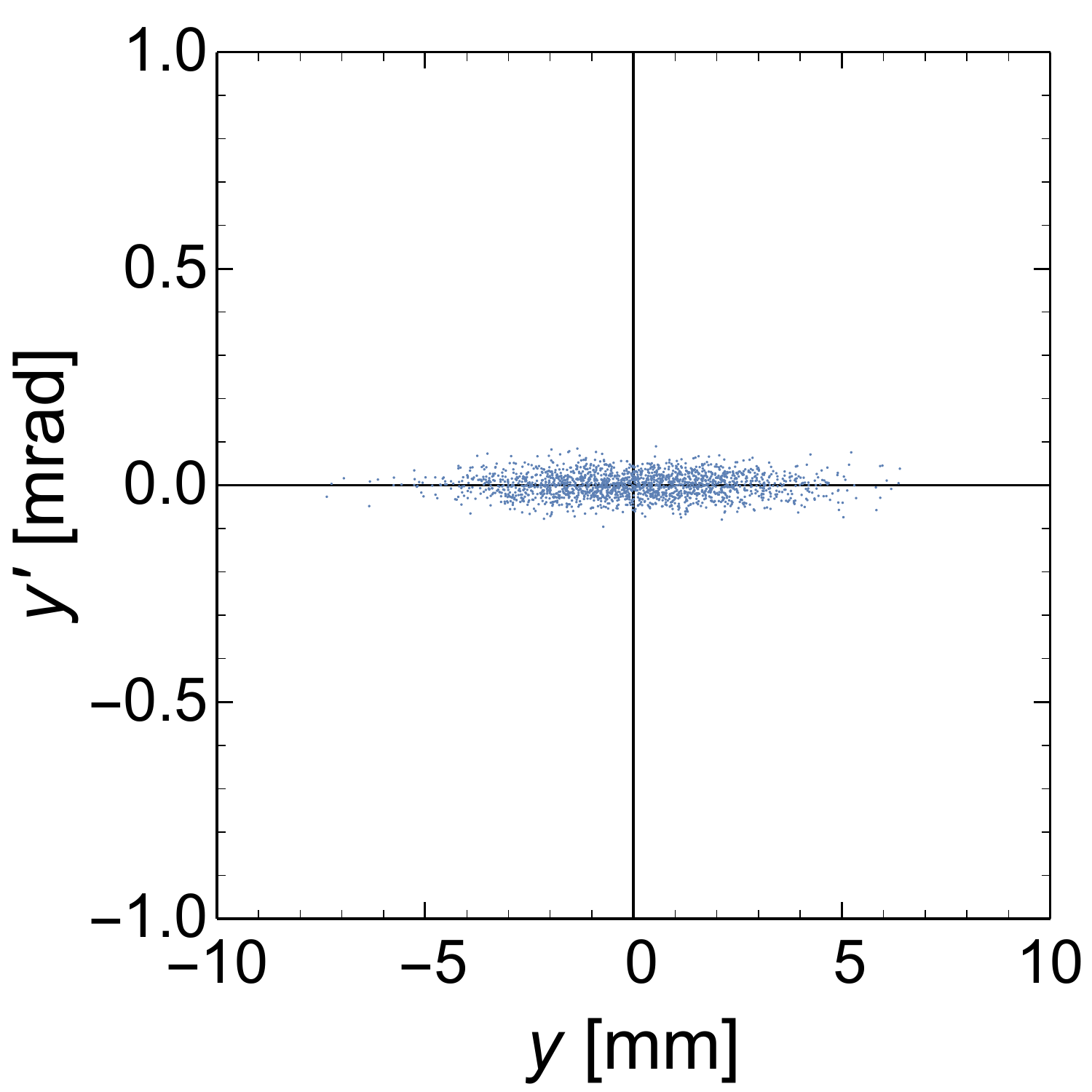}
\caption{
Phase space distributions horizontally, left, and vertically, right, for $\Delta p/p =\pm 0.001$ at the SCCr.  The energy spread of the positron beam was assumed to have an rectangular profile.
}
\label{phase_space_delta_p} 
\end{figure}

\begin{figure}[]
\centering
\includegraphics*[width=100pt]{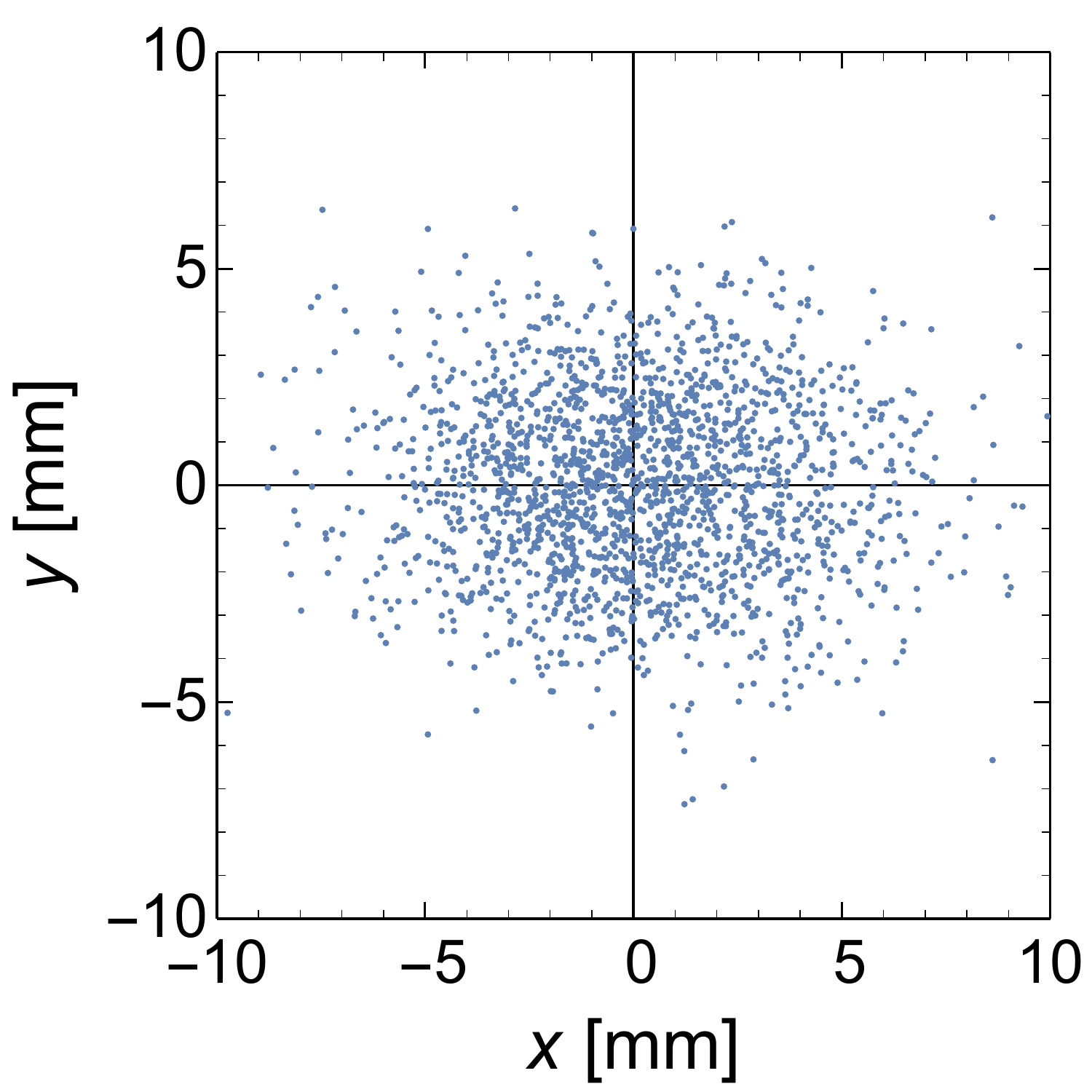}\includegraphics*[width=100pt]{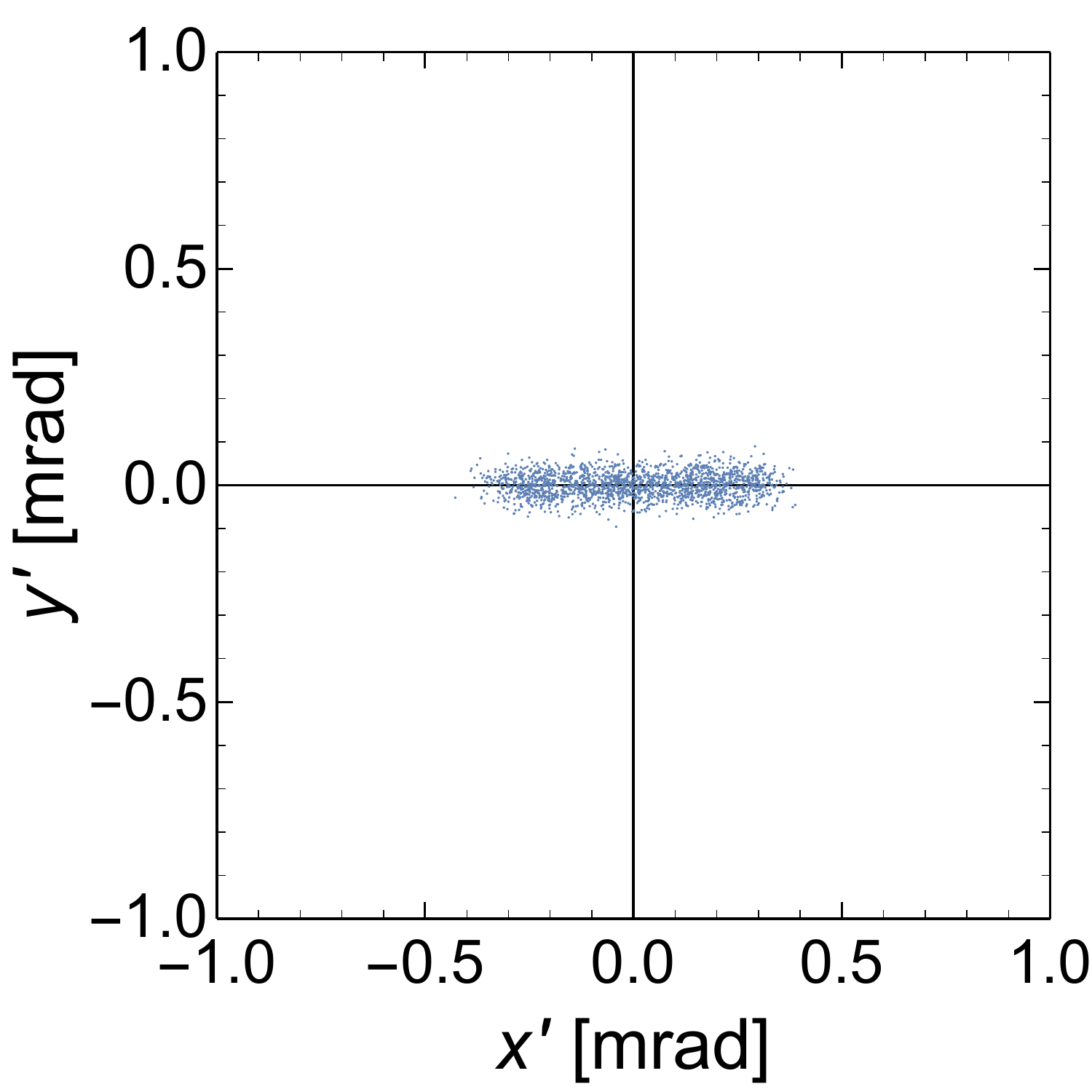}
\caption{
Spot size, left, and angular distribution, right, at the SCCh, for $\Delta p/p = \pm 0.001$. Projected FWHM are $\Delta x$ = 7.7 mm, $\Delta y$ = 5.0 mm, $\Delta x'$ = 0.64 mrad, $\Delta y'$ = 0.064 mrad.
}
\label{spot_size_delta_p} 
\end{figure}

The effect of an accepted energy spread $\Delta T_{+}$ = 1 MeV ($\Delta p/p  = \pm 1.0 \cdot 10^{-3}$) is demonstrated in Fig. \ref{phase_space_delta_p} and Fig. \ref{spot_size_delta_p}. The angular distribution $x'$ in the $(x, x')$ phase space widens while for $y'$ nearly no change in comparison with $\Delta p/p  = 0$ is discernable, compare with Fig. \ref{phase_space} right panel. As demonstrated in Fig. \ref{spot_size_delta_p}, the spread $\Delta T_{+}$ affects only little the beam spot size, however, widens significantly the angular spread $x'$. It turns out that about 82 \% of all events are located within a circle with a radius of 5 mm.

As can be seen from Fig. \ref{spot_size_delta_p}, right panel, the angular distribution $\Delta y'$ = 0.064 mrad is vertically rather narrow. It corresponds to a transverse energy spread of only 1.0 eV.  This feature defines the orientation of the single crystal in the SCCr. The angular spread $\Delta x'$ = 0.35 mrad in $x$ direction is expected to cause low energy tails of the channeling or undulator radiation peaks.

\begin{figure}[h]
\centering
\includegraphics*[width=150pt]{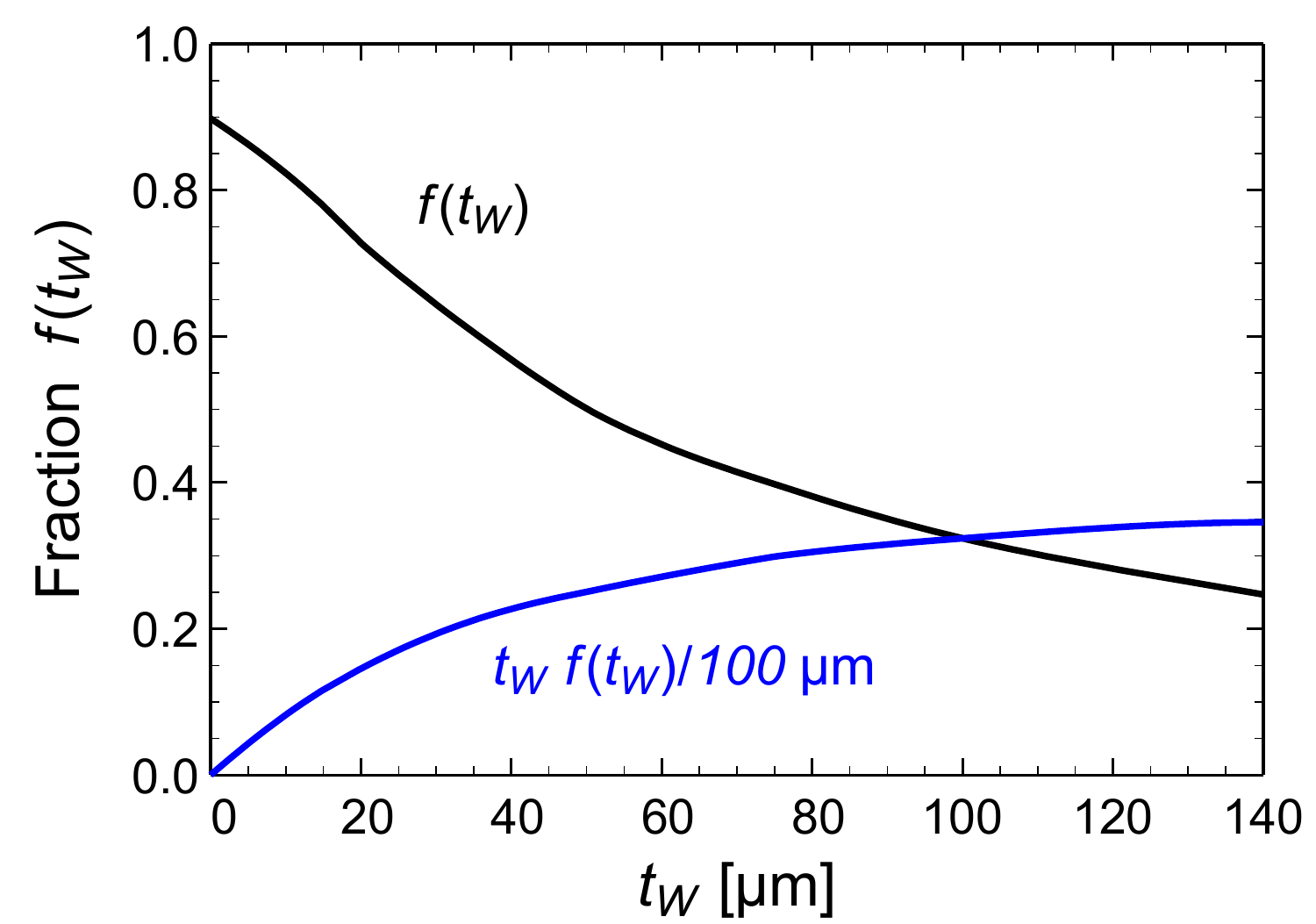}
\caption{
Fraction of positrons accepted within a circle of 5 mm radius at the SCCh position as function of the tungsten self-converter thickness t${}_{w}$ for a positron momentum band $\Delta p/p = \pm 0.001$. Shown is also the product with the target thickness t${}_{W}$.
}
\label{fraction_positrons} 
\end{figure}

\begin{figure}[h]
\centering
\includegraphics*[width=100pt]{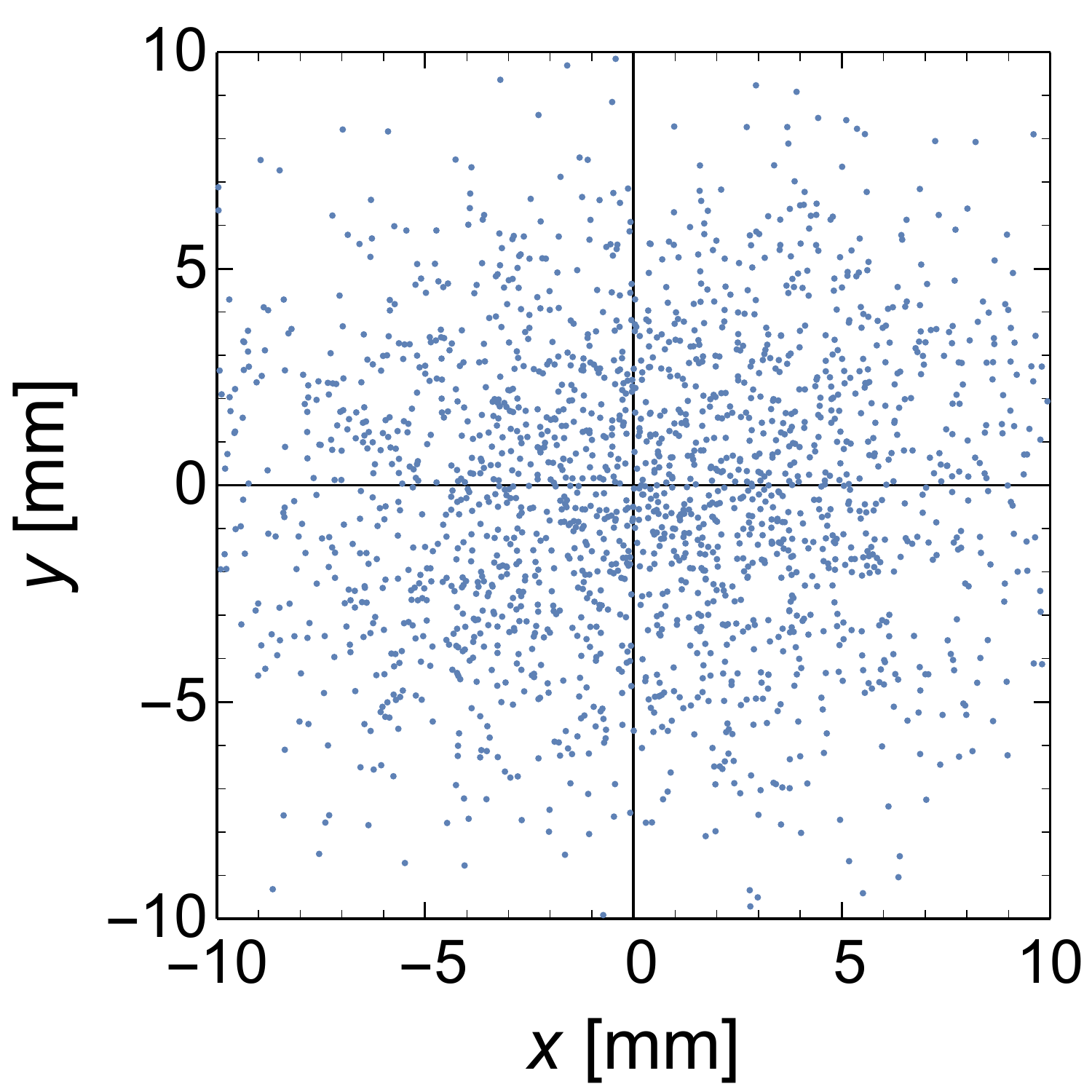}\includegraphics*[width=100pt]{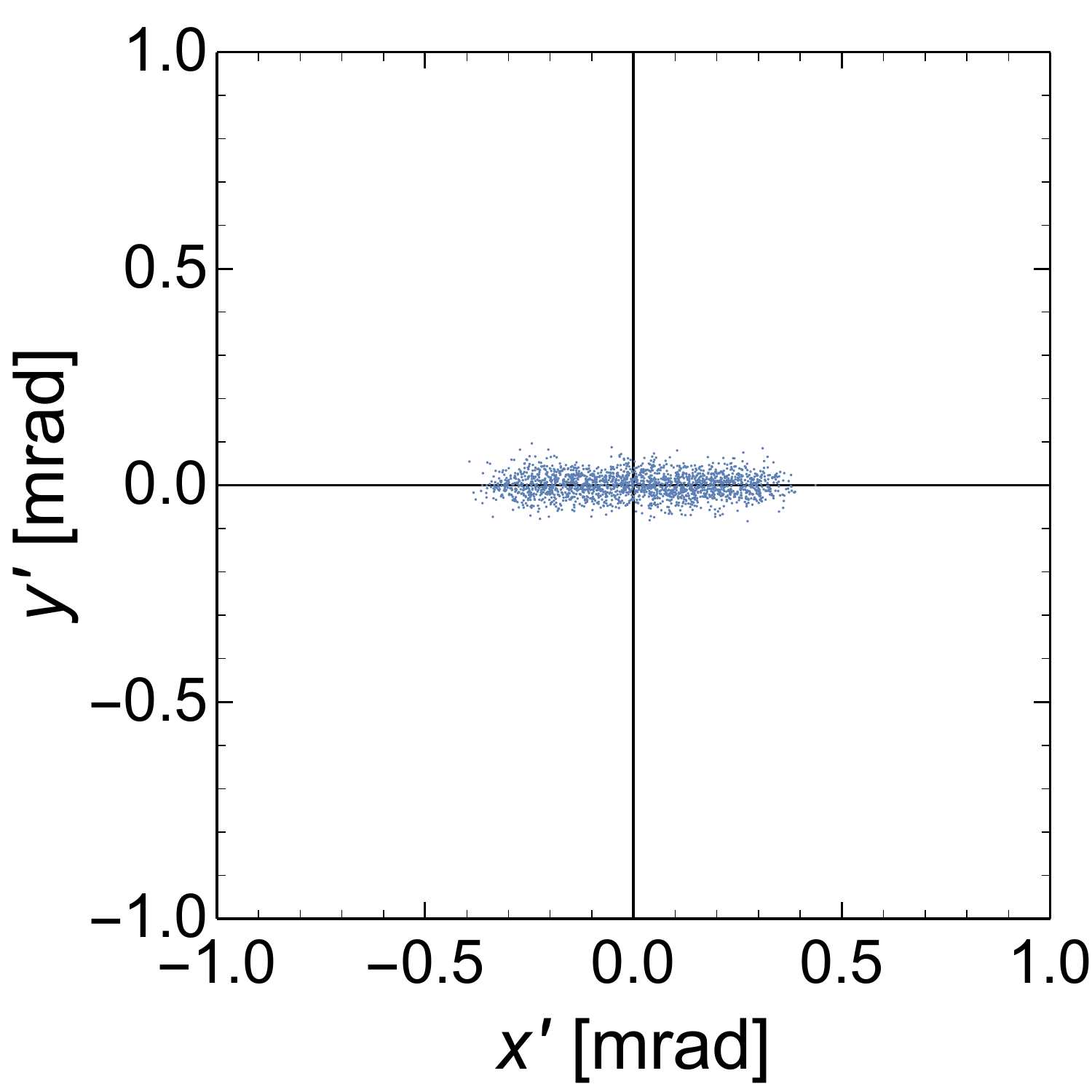}
\caption{
Angular distribution at the SCCh position for a tungsten self-converter thickness t${}_{W}$ = 50 µm. Projected FWHM are $\Delta x$ = 11.7 mm, $\Delta y$ = 8.6 mm, $\Delta x'$ = 0.65 mrad, $\Delta y'$ = 0.066 mrad.
}
\label{characteristicsW50mum} 
\end{figure}

A thickness variation of the tungsten self-converter target has nearly no effect on the scattering distribution at the position of the single crystal chamber SCCh, however, a very significant one on the size of the beam spot. This fact is demonstrated in Fig. \ref{fraction_positrons}. It can be concluded from the products \textit{t${}_{W}$} \textit{f}(\textit{t${}_{W}$}) that the tungsten target thickness can be, in principle, increased by a factor of about 5, resulting in thicknesses of 50 µm, still maintaining the angular distribution shown at Fig. \ref{characteristicsW50mum}. The positron yield scales quadratically with the W-target thickness, resulting in a significant increase of the intensity.

\subsection{Separated-converter geometry}
In the separated-converter geometry, see Fig. \ref{positronYieldWPbInset}, the positron converter target is located also inside the bending magnet BM1, however, not within the electron beam. The bremsstrahlung target should be positioned as close as possible to the positron converter target, preferably outside the bending magnet BM1 since parasitic positrons should be excluded from an acceptance by the beamline. Assuming as a safe distance between the electron beam and the positron converter target 12.4 mm, the Pb positron converter target may be located at $z_{L,Pb}$ = 196 mm, similar as for the self-converter geometry. In principle, all the calculation described in the previous subsection can be performed also for this geometry. We avoid to present results which look similar to those shown in Figs. \ref{phase_space}, \ref{spot_size}, \ref{phase_space_delta_p}, and \ref{spot_size_delta_p}. However, there are differences which originate from broader scattering distributions due to the thicker lead target with larger charge number, and an increased bremsstrahlung spot size at the positron converter target. These effects must be minimized in a future study to fully utilize the larger intensity for this separated-target geometry in comparison with the self-converter geometry.

\section{Discussion} \label{discussion}
Since the positron intensity scales with the bremsstrahlung converter thickness, in the self-converter geometry quadratically, in the separated-converter geometry linearly, the largest possible thickness should be chosen. Beside possible restrictions already discussed, also the angular acceptance of the electron beam line into the dump is of importance. If tails of the scattering distribution scratch constrictions, a large radiation background may be the consequence. Therefore, geometrical bottlenecks in the beam line must be eliminated. Anyway, even after optimization the separated-converter geometry has at least about a factor of two higher intensity in comparison with the self-converter geometry, probably more, however, on the expense of a broader vertical angular spread $y'$ at the SCCh.

Which spot size at the SCCh may be allowed is a question of the observation geometry for the emitted channeling or undulator radiation. Assuming the photon detector is 5 m away, and that the off-axis observation angle should not be larger than 0.4 mrad, the spot size must have a radius of less than 2 mm. Positrons impinging the single crystal target with larger radii can probably be excluded by an anti-coincidence detector.

A count rate estimate will be performed for the 10 µm W target in the self-converter geometry for moderate angular resolution studies. With an anti-coincidence aperture of 2 mm radius in front of the single crystal target in the SCCh the acceptance may be 20 \%. Assuming for channeling or undulator radiation that $5 \cdot 10^{-6}$/keV photons are emitted per 1 keV bandwidth and that the peaks have a half width of 100 keV, the count rate would be $5 \cdot 10^{-6}$/keV $\cdot$ 100~keV $\cdot$ 0.2 $ \cdot$ 13.1/s = 0.0013/s for a positron beam with one MeV bandwidth and 1 nA electron beam current. For reasonable count rates, both, the beam current and the target thickness should be increased. At a beam current of 0.5 $\mu$A, and a somewhat thicker target an increase of the count rate by about a factor $10^3$ should be feasible.

For a high angular resolution study in the self-converter geometry the count rate is significantly lower.

\section{Conclusions} \label{conclusions}
Positron yields have been calculated for a 10 µm thick tungsten target in the self-converter geometry, and for the separated-converter geometry for also a 10 µm thick tungsten bremsstrahlung converter and a 20 µm thick lead positron converter target. The yields at a positron beam energy of 500 MeV, defined as positrons per second, 1 MeV positron beam energy interval and 1 nA electron beam current, are 13.1/(s MeV nA) and 32.8/(s MeV nA) for the self-converter and the separated-converter geometries, respectively. An existing outside open electron beam line bending magnet behaves for positrons like a horizontally focussing and vertically de-focusing quadrupole for which the transfer matrix was calculated. Experimental results are in good agreement with expectations of the positron yield as function of the energy. A beam line has been designed which features in vertical direction an angular divergence of 64 $\mu$rad (FWHM) at a spot size of 5.0 mm (FWHM). At an currently allowed electron beam current of 0.5 $\mu$A, and a tungsten target thickness of 10 $\mu$m, about 6$\cdot10^3$ positrons/s with a band width of 1 MeV could be available for experiments at single crystals.

\section*{Acknowledgements} \label{Acknowledgements}
We acknowledge fruitful discussions with K.-H. Kaiser, M. Negrazus, and Jürgen Ahrens, and their support in the early stage of the experiment.

\section*{Declarations}
This project has been partially supported by the European Commission through the N-LIGHT Project, GA
872196.

This publication is part of a project that has received funding from the European Union’s Horizon 2020 research and innovation programme under grant agreement STRONG – 2020 - No 824093.

\section*{Appendix}
\section*{Parameters of the beam line} \label{appendix A}
According to Eq. (\ref{EQ__13_}) the total transfer matrix is\\
$M_{tot} = S(s_{4} )\cdot M_{BM2}^{PE}(s_{BM2}, R_{BM2},\psi_{PE}) \cdot S(s_{3} )\cdot
Q_{F} (k_{Q2} ,s_{Q2} )\; \cdot S(s_{2} ) \cdot Q_{D} (k_{Q1} ,s_{Q1} )\cdot
S(s_{1} ) \cdot M_{BM1}$.\\
The pole edge focusing has been included in $M_{BM2}^{PE}(s_{BM2}, R_{BM2},\psi_{PE})=M_{PE}(\psi_{PK}, R_{BM2})\cdot M_{BM2}(s_{BM2}, R_{BM2})\cdot M_{PE}(\psi_{PE}, R_{BM2})$.

The following parameters were used:
$s_1$ = 329.0 mm,
$k_{Q1}$ = 3.971$\cdot 10^{-6}$,
$s_{Q1}$ = 300.0 mm,
$s_2$ = 200.0 mm,
$k_{Q2}$ = -2.150$\cdot 10^{-6}$,
$s_{Q2}$ = 300.0 mm,
$s_3$ = 425.0 mm,
$R_{BM2}$ = 2382.6 mm,
$s_{BM2}$ = 652.1 mm,
$\psi_{PE}$= -7.13°,
$s_4$ = 2600.0 mm.
The total transfer matrix reads
\\\\
$M_{tot}=
\\\\
 \left( \small
\begin{array}{ccccc}
 -3.20231 & 2535.2 & 0. & 0. & 2276.2 \\
 -0.000388 & -0.00486 & 0. & 0. & 0.3201 \\
 0. & 0. & -0.593413 & 1895.1 & 0. \\
 0. & 0. & -0.000527 & -0.00129 & 0. \\
 0. & 0. & 0. & 0. & 1. \\
\end{array}
\right)$
\\

At the maximum induction of the second dipole magnet $B_{BM2}$ = 0.7 Tesla the 500 MeV positron beam exits BM2 with an angle of -2.67°.

For the calculation of the positron beam characteristics in the SCCh various spreads at the target position must be known. For the beam spot size $\sigma$${}_{x}$ = 0.1 mm, $\sigma$${}_{y}$ = 0.05 mm and corresponding angular spreads assuming horizontally and vertically emittances $\epsilon_{x} = 10 \cdot 10^{-6} \pi$~mm~rad and $\epsilon_{y} = 2 \cdot 10^{-6} \pi$~mm~rad, respectively, were assumed. The angular distribution for the bremsstrahlung photons has been approximated by a Gaussian with $\sigma$${}_{bs}$ = 0.8/$\gamma$${}_{855}$, i.e., long tails have been neglegted which causes errors of about 7 \% in the intensity. The angular distribution of the converted positrons was assumed to be $\sigma_{+}$ = 0.8/$\gamma_{+}$. The scattering distributions of the 855 MeV electrons and 500 MeV positrons in the W target were assumed for the 10 $\mu$m thick W target to be in the mean $\sigma_{sc}$(855 MeV)/2 = 0.33 mrad, $\sigma_{sc}$(500 MeV)/2 = 0.57 mrad. Where appropriate, distributions were added quadratically.

\section*{Data availability statement} 
All data are freely available.

\section*{Author contribution statement}
H. Backe was responsible for all calculations, with exception of the magnetic field of BM1 which was done by P. Heil, and text editing. W. Lauth conducted the experiment and was responsible also for text editing. F. Stieler constructed the calorimeter depicted in Fig. \ref{lyso}. B. Ledroit performed measurements of the magnetic field and made calculations for BM1. P. Drexler and P. Klag were responsible for data acquisition and data analysis.

\bibliographystyle{spphys}
\bibliography{bibfileWL}



\end{document}